\documentclass[traditabstract]{aa} % for the abstract without structuration 
\usepackage{longtable}
\usepackage{latexsym}
\usepackage{amssymb}
\usepackage{lscape}
\usepackage[]{natbib}

\usepackage{color}

\usepackage{graphicx}
\usepackage{txfonts}
\def\lsim{\mathrel{\rlap{\lower 3pt \hbox{$\sim$}} \raise 2.0pt \hbox{$<$}}}
\def\gsim{\mathrel{\rlap{\lower 3pt \hbox{$\sim$}} \raise 2.0pt \hbox{$>$}}}

\title{The role of bars in quenching star formation from $z$ = 3 to the present epoch}
\subtitle{H$\alpha3$: an H$\alpha$ imaging survey of HI selected galaxies from ALFALFA, VI\thanks{Based on observations 
taken at the  observatory of San Pedro Martir (Baja California, Mexico), belonging to the
Mexican Observatorio Astron\'omico Nacional.}}

\author{G. Gavazzi \inst{1}                                 
\and G. Consolandi \inst{1}                                 
\and M. Dotti \inst{1,2}                                    
\and R. Fanali \inst{1}                                     
\and M. Fossati \inst{3,4}                                  
\and M. Fumagalli \inst{5,6}                                
\and E. Viscardi \inst{1}                                   
\and G. Savorgnan \inst{7}                                  
\and A. Boselli \inst{8}                                    
\and L. Guti\'errez \inst{9}                                
\and H. Hern\'andez Toledo \inst{10}                        
\and R. Giovanelli \inst{11}                                
\and M.P. Haynes \inst{11}                                  
}
\authorrunning{G. Gavazzi et al.}
\titlerunning{H$\alpha3$: H$\alpha$ imaging survey of HI selected galaxies from ALFALFA}
\institute{Universit\`a degli Studi di Milano-Bicocca, Piazza della Scienza 3, 20126 Milano, Italy\\
\email {giuseppe.gavazzi@mib.infn.it}
\and
INFN, Sezione di Milano-Bicocca, Piazza della Scienza 3, 20126 Milano, Italy
\and 
Universit{\"a}ts-Sternwarte M{\"u}nchen, Schenierstrasse 1, D-81679 M{\"u}nchen, Germany. 
\and
Max-Planck-Institut f{\"u}r Extraterrestrische Physik, Giessenbachstrasse, D-85748 Garching, Germany\\
\email {mfossati@mpe.mpg.de}
\and
Institute for Computational Cosmology, Department of Physics, Durham University, South Road, Durham, DH1 3LE, UK\\
\email {michele.fumagalli@durham.ac.uk}
\and 
Carnegie Observatories, 813 Santa Barbara Street, Pasadena, CA 91101, USA.
\and
Centre for Astrophysics and Supercomputing, Swinburne University of Technology, Hawthorn, Victoria 3122, Australia\\
\email {gsavorgn@astro.swin.edu.au}
\and
Aix Marseille Universit\'e, CNRS, LAM (Laboratoire d'Astrophysique de Marseille) UMR 7326, 13388, Marseille, France\\
\email {alessandro.boselli@lam.fr}
\and
Instituto de Astronom\'ia, Universidad Nacional Aut\'onoma de M\'exico, 
Carretera Tijuana-Ensenada, km 103, 22860 Ensenada, B.C., M\'exico.\\
\email {leonel@astro.unam.mx}
\and
Instituto de Astronom\'ia, Universidad Nacional Aut\'onoma de M\'exico, 
Apartado Postal 70-264, 04510 M\'exico D.F., M\'exico.\\
\email {hector@astroscu.unam.mx}
\and
Center for Radiophysics and Space Research, Space Science Building, Ithaca, NY, 14853\\
\email {haynes@astro.cornell.edu, riccardo@astro.cornell.edu}
}
\begin{document}

\date{Received; accepted}

% \abstract{}{}{}{}{} 
% 5 {} token are mandatory
 
\abstract {A growing body of evidence indicates that the star formation rate
  per unit stellar mass (sSFR) decreases with increasing mass in normal
  main-sequence star-forming galaxies. Many processes have been advocated 
  as being responsible for this trend (also known as {\it \emph{mass quenching}}), e.g., feedback from 
  active galactic nuclei (AGNs),   and the formation of classical bulges.  In order to 
  improve our insight into the mechanisms 
  regulating the star formation in normal star-forming galaxies  across cosmic epochs, we
  determine a refined  star formation versus stellar mass relation in the local
  Universe.  To this end we use the H$\alpha$ narrow-band imaging follow-up
  survey (H$\alpha$3) of field galaxies selected from the HI Arecibo Legacy
  Fast ALFA Survey (ALFALFA) in the Coma and Local superclusters.  
  By complementing this local determination with high-redshift measurements 
  from the literature, we reconstruct the star formation history of main-sequence galaxies 
  as a function of stellar mass from the present epoch up to $z=3$.  
  In agreement with previous studies, our analysis shows that quenching mechanisms occur 
  above a threshold stellar mass $M_{\rm knee}$ that evolves with redshift as
  $\propto (1+z)^{2}$.  Moreover, visual morphological classification of individual 
  objects in our local sample
  reveals a sharp increase in the fraction of visually classified strong bars with mass, hinting that
  strong bars may contribute to the observed downturn in the sSFR above $M_{\rm knee}$.
  We test this hypothesis using a simple but physically motivated numerical model for bar formation,
  finding that strong bars can rapidly quench star formation in the central few kpc of field galaxies.
  We conclude that strong bars contribute significantly to the red colors observed in the inner parts 
  of massive galaxies,
  although additional mechanisms are likely required to quench the star formation in the outer regions 
  of massive spiral galaxies.
  Intriguingly, when we extrapolate our model to higher redshifts, 
  we successfully recover the observed redshift evolution for $M_{\rm knee}$. 
  %Our study highlights how the formation of strong bars in massive galaxies is
  %an important mechanism in regulating the redshift evolution of the sSFR for field 
  %main-sequence galaxies.
}

\maketitle

%
%________________________________________________________________

\section{Introduction}

 Unlike starburst galaxies,
 normal star-forming galaxies 
inhabit the main sequence at all redshifts (e.g., Noeske et al. 2007, Elbaz et
al. 2011).  Among local main-sequence galaxies, the dependence of the star
formation rate on the stellar mass is still debated in the literature.  In
other words, it has  not yet been determined whether  
the specific star formation rate (sSFR) decreases with increasing 
 stellar mass (a process also known as {\it \emph{mass quenching}} or
 {\it \emph{downsizing}}, Cowie et al. 1996; Gavazzi et al. 1996; Boselli et al. 2001, Fontanot et
 al. 2009; Gavazzi 2009; Huang et al. 2012) or whether these 
 two quantities are nearly proportional at all masses (e.g., Peng et al. 2010).
A broader consensus exists  instead on the quenching of massive 
main-sequence galaxies at higher redshift, where massive galaxies 
are seen to evolve more rapidly (e.g., Whitaker
et al. 2014; Ilbert et al. 2014) than their less-massive counterparts. 
However, some tension remains between the observations and the current models and
simulations of galaxy evolution (Fontanot et al. 2009; Weinmann et al. 2009, 2012; 
Henriques et al. 2013; Boylan-Kolchin et al. 2012; 
Hirschmann et al. 2014) emphasizing that the
physics of the quenching of star formation is still not fully understood.
The nature of the physical processes responsible for 
this mass quenching is still under debate (Peng et al. 2012, Lilly et al. 2013).  

Several mechanisms are often invoked, including AGN feedback
  (e.g., Scannapieco et al. 2005; Bundy et al. 2008; Oppenheimer et
  al. 2010; Tessier et al. 2011); cosmological starvation (e.g., Feldmann \& Mayer
  2015; Fiacconi et al. 2015); and formation of kinematically hot
  spheroidal structures such as classical bulges, which are thought to form
  through rapid merger events (e.g., Aguerri et al. 2001) or multiple coalescence of
  giant clumps in primordial disks (e.g., Elmegreen et al. 2008). The final word on
  the relative importance of these (or other) quenching processes has not been
  spoken yet.

In  this paper, starting from Sect.~\ref{z0}, we exploit the recently completed
H$\alpha$3 survey in the Coma Supercluster (see the accompanying Paper V of
this series; Gavazzi et al. 2015) and in the Local Supercluster 
(Gavazzi et al. 2012, Paper I) to add a further piece of evidence 
in support of a significant quenching of star formation at masses 
$M_* > M_{\rm knee} \approx 10^{9.5}$ M$_{\odot}$ for local, normal late-type galaxies. 
In Section~\ref{highz} we also show that the threshold mass $M_{\rm knee}$
for the quenching increases with redshift. 
By exploiting the low redshift nature of our sample for which visual 
morphological classification can be obtained, we show that the occupation fraction of
visually classified strong bars 
drops drastically for $M_* \lsim M_{\rm knee}$ (Section \ref{bars}).
With the aid of numerical and analytical arguments, in section~\ref{theory},
we develop a simple, observationally driven argument to explain the existence of a 
threshold mass for the formation of strong bars,  which in turn contributes to the observed 
quenching. This model also predicts the observed 
redshift-dependence of $M_{\rm knee}$. Discussion and conclusions follow 
in sections \ref{speculations} and ~\ref{discussion}.

\section{Star formation rate at $z$=0}
\label{z0}

The sample of star-forming galaxies at $z$=0 used in this work consists of
1399 galaxies HI-selected primarily from ALFALFA (Haynes et al. 2011)  in the regions of
the Local Supercluster and in the Coma Supercluster. 
These are complemented with pointed HI
observations of late-type galaxies taken at similar sensitivity in the region of the Coma supercluster not covered by ALFALFA
(as listed in the GOLDMine database of Gavazzi et al. 2003, 2014).
Gavazzi et al. (2008,
2013a) showed that ALFALFA selected galaxies are genuine star-forming objects
(late-type galaxies, LTGs) with almost no contamination from S0s and S0as (see
also Buat et al. 2014 for a discussion on the selection criteria of star-forming galaxies).  Among these,  1091 were followed up with H$\alpha$ imaging
observations to derive their global star formation rates (SFRs; Gavazzi et al 2015,
Paper V).  The H$\alpha$ luminosity was corrected for Galactic extinction,
deblending from [NII], and  internal extinction following Lee et
al. (2009).  
Throughout this series (including  Paper V), stellar masses $M_*$ and SFRs have been 
computed assuming a Salpeter Initial Mass Function (IMF), following the calibrations of Kennicutt (1998). 
In this paper, however, we compare results from our survey with literature values.  
We therefore recompute both stellar masses and SFR assuming a Chabrier IMF, 
as commonly done in the modern literature. Specifically, the transformations applied to 
the H$\alpha$3 survey are $ \rm SFR_{Chabrier}= 1.5 \times SFR_{Salpeter}$ and 
$\log (M_*/{\rm M_\odot)}=-0.963+1.032 ~(g-i)+\log(L_i/\rm L_\odot)$, following 
Zibetti et al. (2009).

The H$\alpha$3 survey also includes  galaxies in proximity and inside the rich 
Coma and Virgo clusters. The present study  focuses on unperturbed 
galaxies, which we select to avoid environmental quenching effects (see Gavazzi et al. 2013b).
To this purpose, we do not include in our analysis galaxies with HI-deficiency parameters greater than
0.3 \footnote{The HI deficiency parameter, defined by Haynes \& Giovanelli
(1984) provides the logarithmic difference between the HI mass actually
observed in galaxies and the one inferred from their optical diameter.}. 

\begin{figure*}[!t]
\centering
\includegraphics[width=9cm, trim = 0cm 0cm 0cm 0cm]{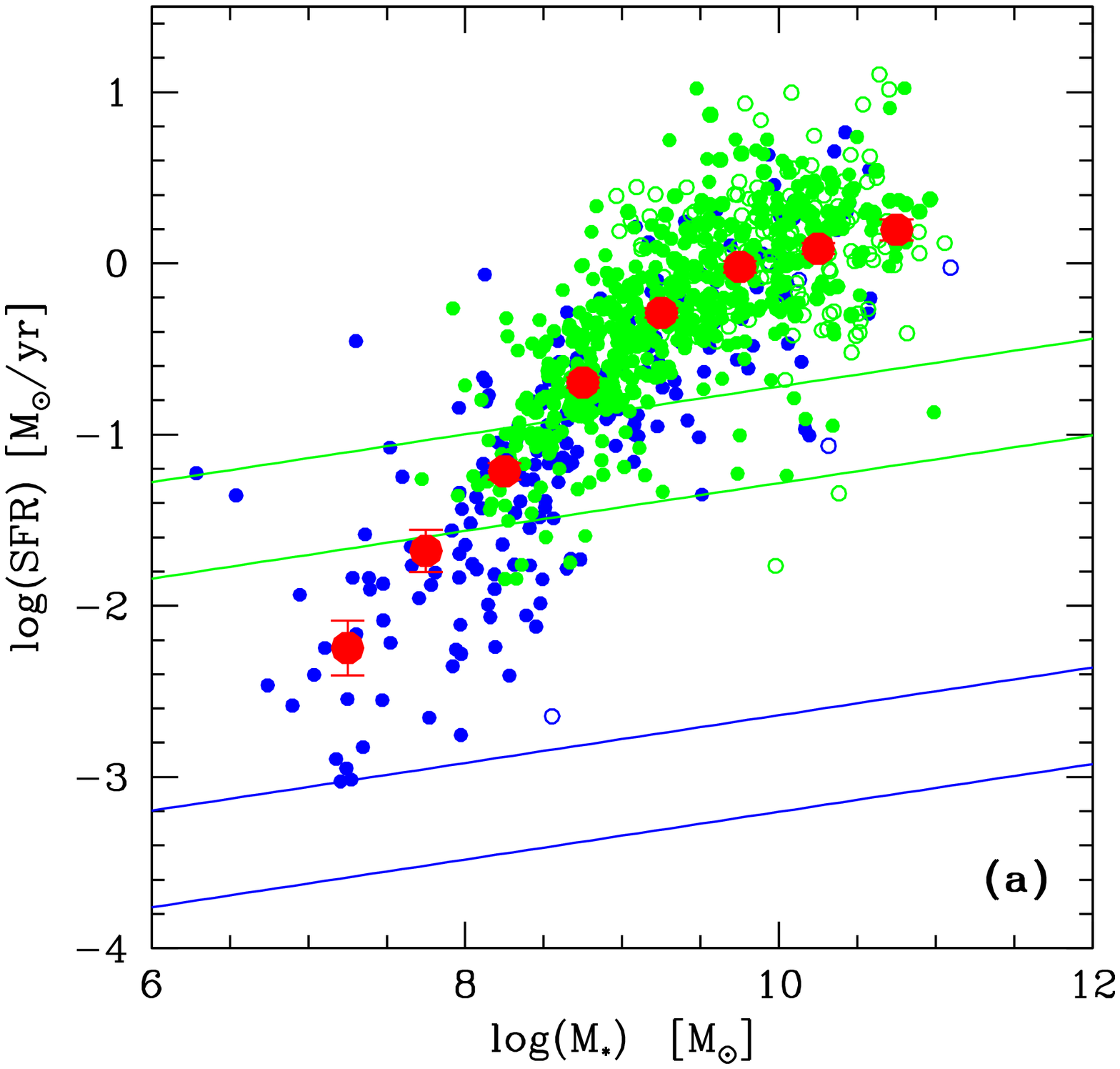}
\includegraphics[width=9cm, trim = 0cm 0cm 0cm 0cm]{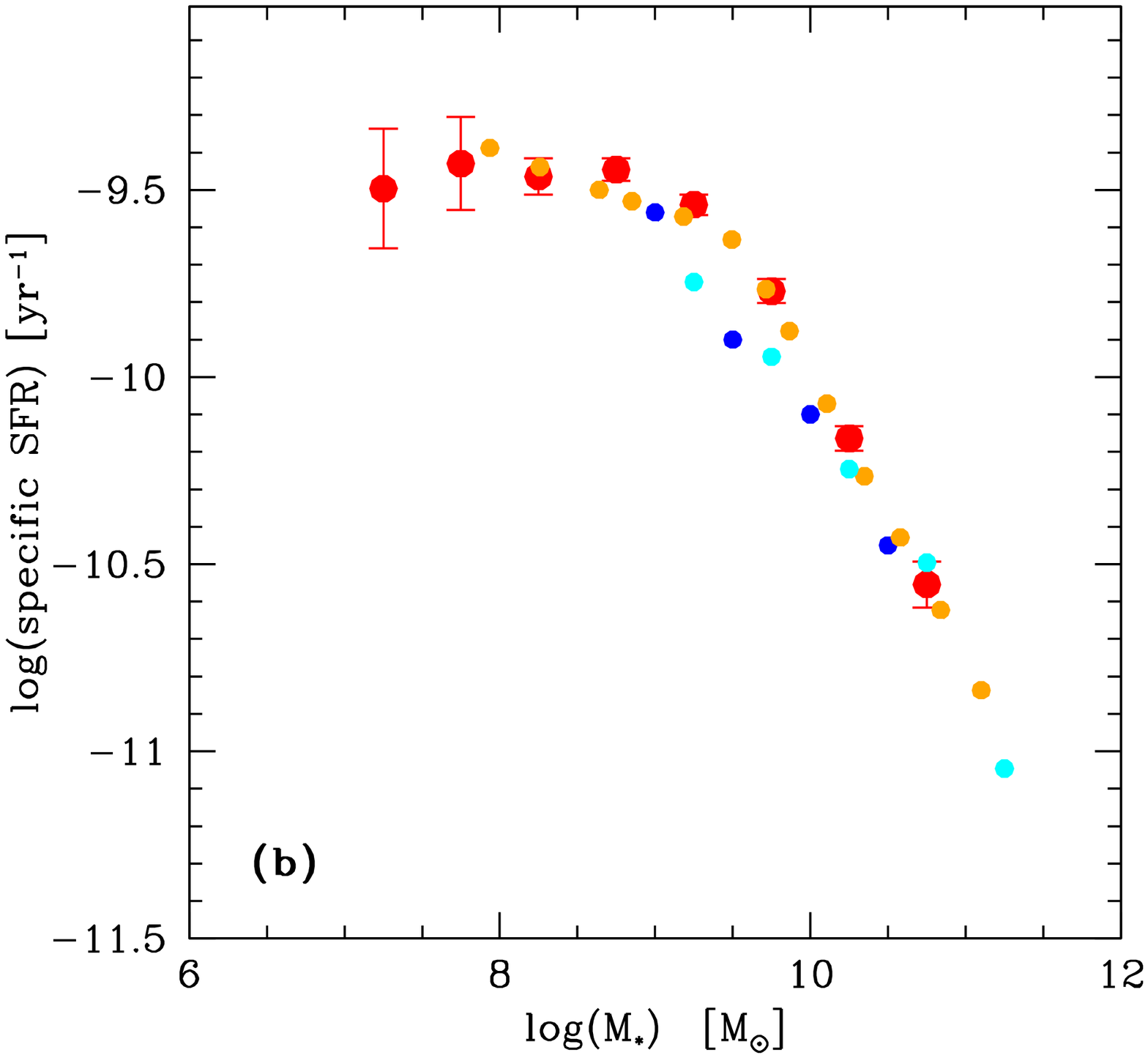}
\caption{(Panel a) The star formation rate as a function of stellar mass at
  $z$=0 for HI non-deficient galaxies. Green symbols represent galaxies  
  in the Coma supercluster; blue symbols are in the Local
  Supercluster. Red symbols are averages in bins of stellar mass.  The derived
  star formation rate are computed from the H$\alpha$ luminosity assuming a
  Chabrier IMF.  The two green (blue) diagonal lines represent the  selection bias on the
  SFR induced by the limited sensitivity of ALFALFA at the distance of Coma
  (Virgo), computed for galaxies with inclination of 10 and 45 degrees
  respectively.
  (Panel b): the specific star formation rate  as a
  function of stellar mass at $z$=0. 
  Average values from our local sample
  (Coma+Virgo) are given with red dots with error bars. Orange points are from Huang et al. (2012) and
  cyan points are from   Brinchmann et al. (2004) (SDSS at $z$=0). 
  The blue points are taken in the interval $0.05<z<0.08$ from Bauer et al. (2013).
  All sets of points show remarkable
  consistency above $\rm 10^{9.5}$ M$_\odot$.  
  }
 \label{sfr0}  
\end{figure*}

In addition to  the cut based on HI deficiency, we wish to remove any possible residual environmental effects, such as sSFR 
quenching in high-density environments (e.g., Poggianti at al. 1999; Lewis et al. 2002; Balogh et al. 2004; Patel et
al. 2009; Boselli \& Gavazzi 2006, 2014). 
Following Gavazzi et al. (2010) we measured around all galaxies (in the Local and Coma superclusters irrespective 
of their type and HI content) a density contrast 
$\delta_{1,1000}$, computed within a cylinder of 1 $h^{-1}$Mpc radius and 1000 $\rm km s^{-1}$  half-length.
We repeated the analysis shown in Figure \ref{sfr0} by including only galaxies with $\delta_{1,1000} < 20$, 
this time avoiding the cores of the rich clusters Virgo, A1367 and Coma. 
Except for a marginal decrease in the number of objects below $M_* = \rm 10^{8.5}$ M$_\odot$, no differences are seen 
at high mass that could explain the observed decrease of the sSFR as being due to environmental mechanisms.  

With this selection, and combining two local samples in the Local and Coma superclusters, 
we obtain a final sample of  864 galaxies.  The derived
star formation rates are plotted in Figure \ref{sfr0}(a) and listed in 
Table \ref{tab1} as a function of stellar mass.

The flux limit of ALFALFA translates into a 
selection effect in the HI mass, which depends on galaxy inclination (see Giovanelli et al. 2005). 
At the distance of Virgo, this limit is log$(M_{\rm HI}/$M$_\odot)=7.25-7.54$, computed for
inclinations of 10 and 45 degrees, respectively. 
As discussed in Gavazzi et al. (2015), this selection threshold does not hamper the 
detection of normal gas-rich galaxies with typical stellar masses as low as $10^7$M$_\odot$.
This sensitivity limit is, however, 25 times worse at the distance of Coma,
being log$(M_{\rm HI}/$M$_\odot)=8.78-9.08$.  Owing to this shallower selection, only
an incomplete set of LTG galaxies at the distance of Coma are detected by
ALFALFA and have been followed up by H$\alpha$3.
The galaxies included in our study are therefore the most HI-rich objects, 
which means that the corresponding star formation rates are generally biased towards 
high values. The two diagonal lines in Figure
\ref{sfr0}(a) show this selection effect for Coma  and for Virgo.
  Because of this bias, the slope of the SFR versus mass relation  is
significantly flatter for Coma than for the Local supercluster.  
Conversely, one can note how this latter subsample is not hampered by the ALFALFA 
selection bias, but it suffers instead from an undersampling at the highest mass bin, owing to a lack of surveyed
volume. However, the two subsamples are complementary, and the underlying SFR versus mass
relation can be obtained by combining them together. The mean SFRs in bins of stellar mass 
for this combined sample is shown in Figure \ref{sfr0}(a).  Here, we also show that the star
formation rate of star-forming galaxies (main-sequence galaxies) in the local Universe
is inconsistent with a single power law (a slope of nearly unity), but shows a
decreasing slope with increasing mass.

Figure~\ref{sfr0}(b) shows the specific star formation rate derived from 
our data. 
Another set of local HI-selected galaxies in the entire ALFALFA survey by
Huang et al. (2012) is shown. Although it is derived with
a different SFR indicator based on UV luminosity corrected for IR, this second sample 
is remarkably consistent with our data.
Additionally, we show a third sample of star-forming galaxies  from 
Brinchmann et al. (2004), derived in the local Universe using SDSS data corrected for aperture effects.
Finally we show  a set of local data (obtained at $0.05<z<0.08$) from the GAMA survey by Bauer et al. (2013).
Despite the different selections and indicators, all local determinations are in reasonable mutual agreement. 
Although not shown in Figure \ref{sfr0}(b), we note that 
the SFR versus stellar mass relation derived by Peng et al. (2010) using SDSS data is inconsistent 
with that found in other local samples, mainly because it does not show a change of slope
above some turnover mass. We think this is due to the choice of Peng et al. (2010) to
restrict their star-forming sample to galaxies showing strong emission lines
in the nuclear spectra, thus biasing the selection towards starbursting objects.
Similar inconsistency with Peng et al. (2010) is reported in Bauer et
al. (2013) in their determination of the local SFR from the GAMA survey.
%{\mfo io qui metterei anche i punti di GAMA nel redshift bin piu basso. E nella figura 
%2 quelli di GAMA nel redshift bin piu' alto. Poi se non vanno deep fino al turnover mass non li usiamo 
%in Figura3 ma almeno riempiamo il piu possibile le figure 1 e 2 con i dati della lettaratura.
%Mi sto chiedendo anche perche' non mettere Peng nella figura 1 visto che lo si cita.

\begin{table}[t]
\caption{Star formation sequence at $z$=0.   
The associated uncertainties are Poissonian.}
\centering
\begin{tabular}{c c c }
\hline\hline
log $M_*$ bin         &   logSFR  &  Error   \\
     M$_\odot$         &     M$_\odot$~yr$^{-1}$          &      M$_\odot$~yr$^{-1}$        \\
\hline\hline
 7.0   -  7.5   &   -2.247  &  0.159  \\
 7.5   -  8.0   &   -1.680  &  0.124  \\
 8.0   -  8.5   &   -1.214  &  0.047  \\
 8.5   -  9.0   &   -0.696  &  0.030  \\
 9.0   -  9.5   &   -0.290  &  0.027  \\
 9.5   -  10.0  &   -0.021  &  0.032  \\
 10.0  -  10.5  &    0.086  &  0.032  \\
 10.5  -  11.0  &    0.196  &  0.061  \\
 \hline
\end{tabular}
\label{tab1}
\end{table}

\section{The star formation rate as a function of redshift}
\label{highz}

\begin{figure}[!t]                                                                                            
\centering                                                                                                    
\includegraphics[width=9cm, trim = 0cm 0cm 0cm 0cm]{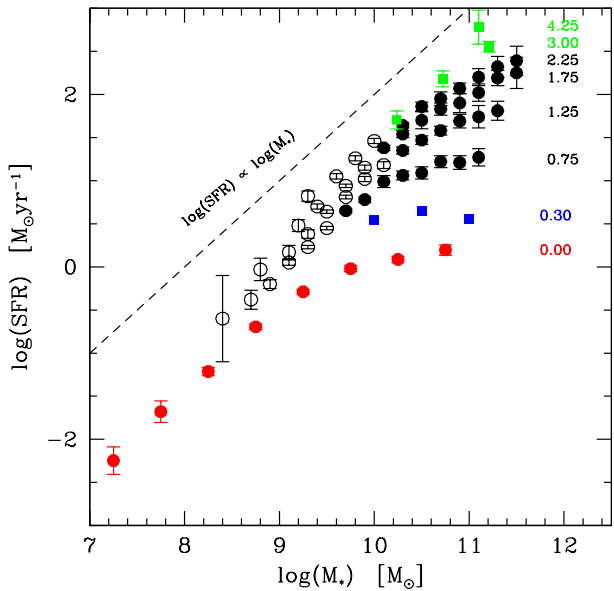}                                                
\caption{The star formation rate as a function of stellar mass in bins  of redshift.                             
 Data at $z$=0 (red) are from this work (red symbols in Figure \ref{sfr0}).
 Data at $z$=0.3 (blue) are from Bauer et al. (2013).
 Measurements in the interval $0.75<z<2.25$ (black) are from Whitaker et al. (2104)
 (empty circles are for mass bins where individual galaxies were stacked when deriving IR luminosities);                       
 the points at $z$=3 and $z$=4.25 (green) are from Schreiber et al. (2015). Whitaker et al. (2014) and Schreiber et al. (2015)
 data are plotted above their respective completeness limit.}                                 
\label{sfr}                                                                                                   
\end{figure}                                                                                                  

In this section we extend the analysis to the star formation rate from $z$=0
up to $z \sim 4$. Figure \ref{sfr} gives the SFR as a function of stellar
mass in bins of increasing $z$.  The local data from this work (red) are taken
from Figure \ref{sfr0}. 
Data at $z$=0.3  are from the GAMA survey by Bauer et al. (2013).
Data in the 0.75, 1.25, 1.75, 2.25 redshift bins  are
from Whitaker et al. (2014), who selected star-forming galaxies using the UVJ
diagram (Williams et al. 2009). Their  SFR  are derived combining UV and IR luminosities
from the deep CANDELS+3DHST surveys (Skelton et al. 2014)
to account  for obscured and unobscured star formation. 
This is currently among the best indicators of star formation at high-$z$ (Wuyts et al. 2011).
At even higher redshift ($z$=3 and $z$=4.25) we show the
recent measurements by Schreiber et al. (2015), who adopt the same SFR indicator 
computed using FIR Herschel calibrated SFRs complemented by the UV
luminosity from SED fitting.
A line of proportionality between SFR and mass (exponential stellar mass
growth) is given to guide the eye.

\begin{figure*}[!t]
\centering
\includegraphics[width=9cm, trim = 0cm 0cm 0cm 0cm]{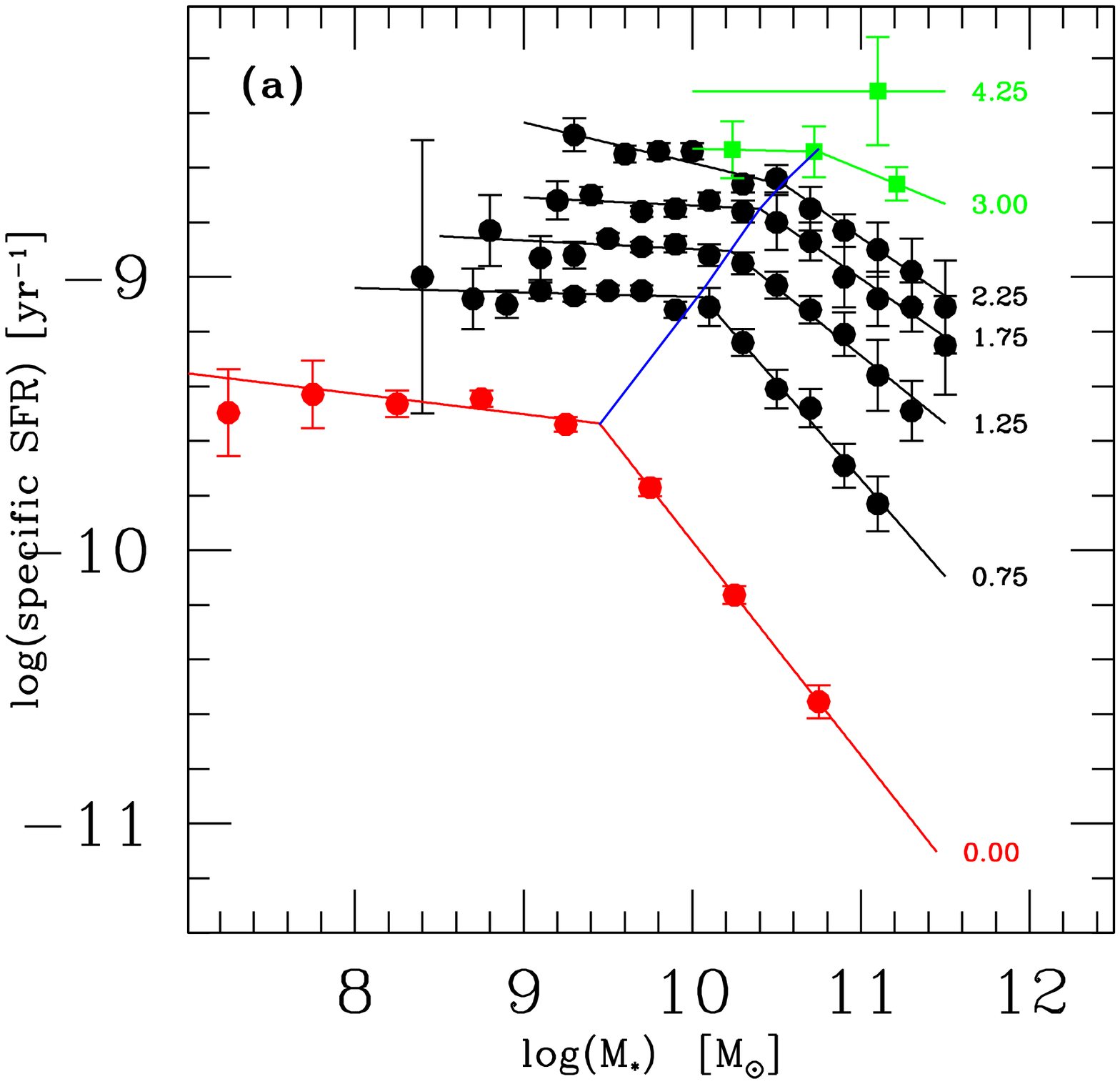}
\includegraphics[width=9cm, trim = 0cm 0cm 0cm 0cm]{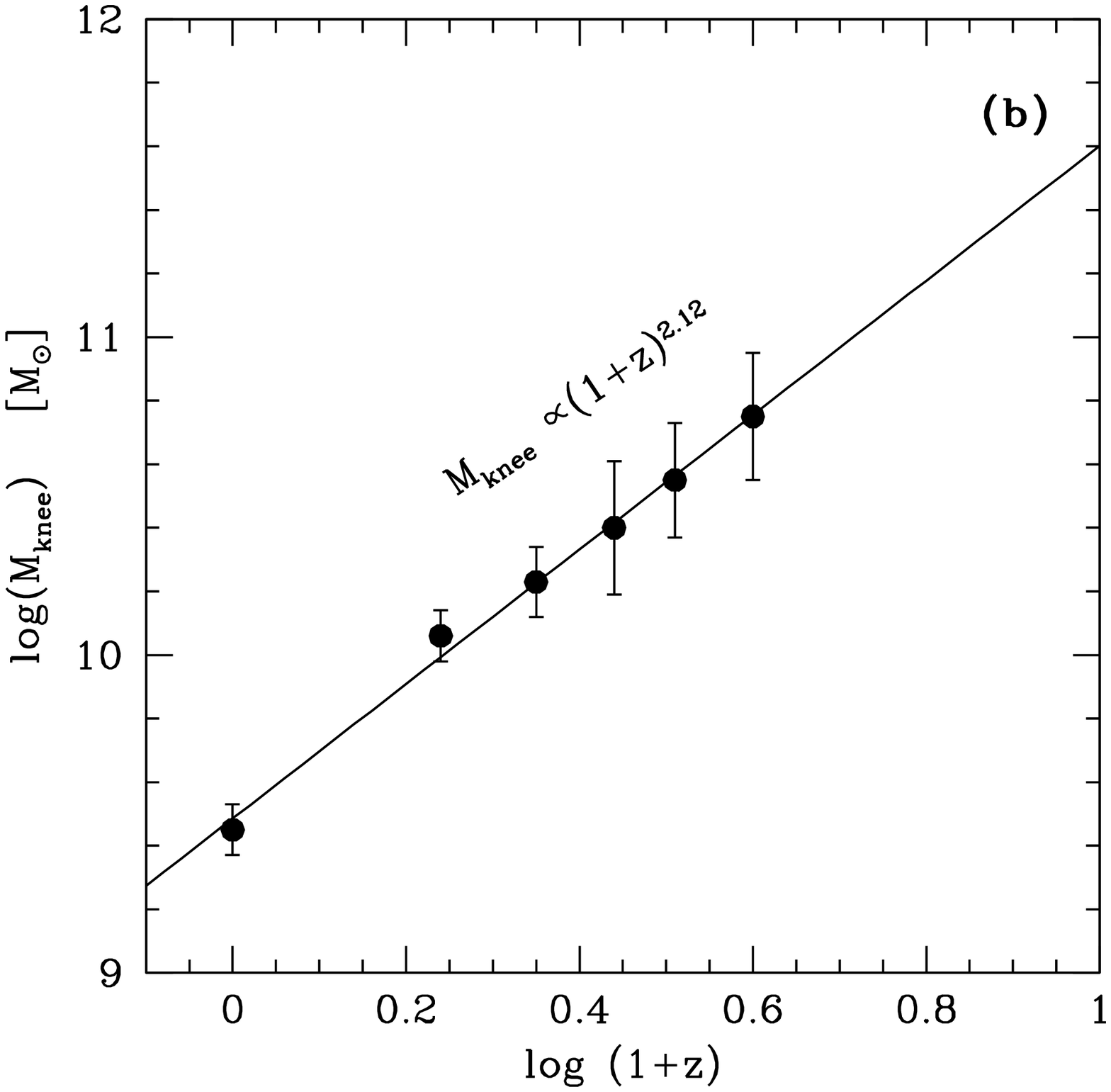}
\includegraphics[width=9cm, trim = 0cm 0cm 0cm 0cm]{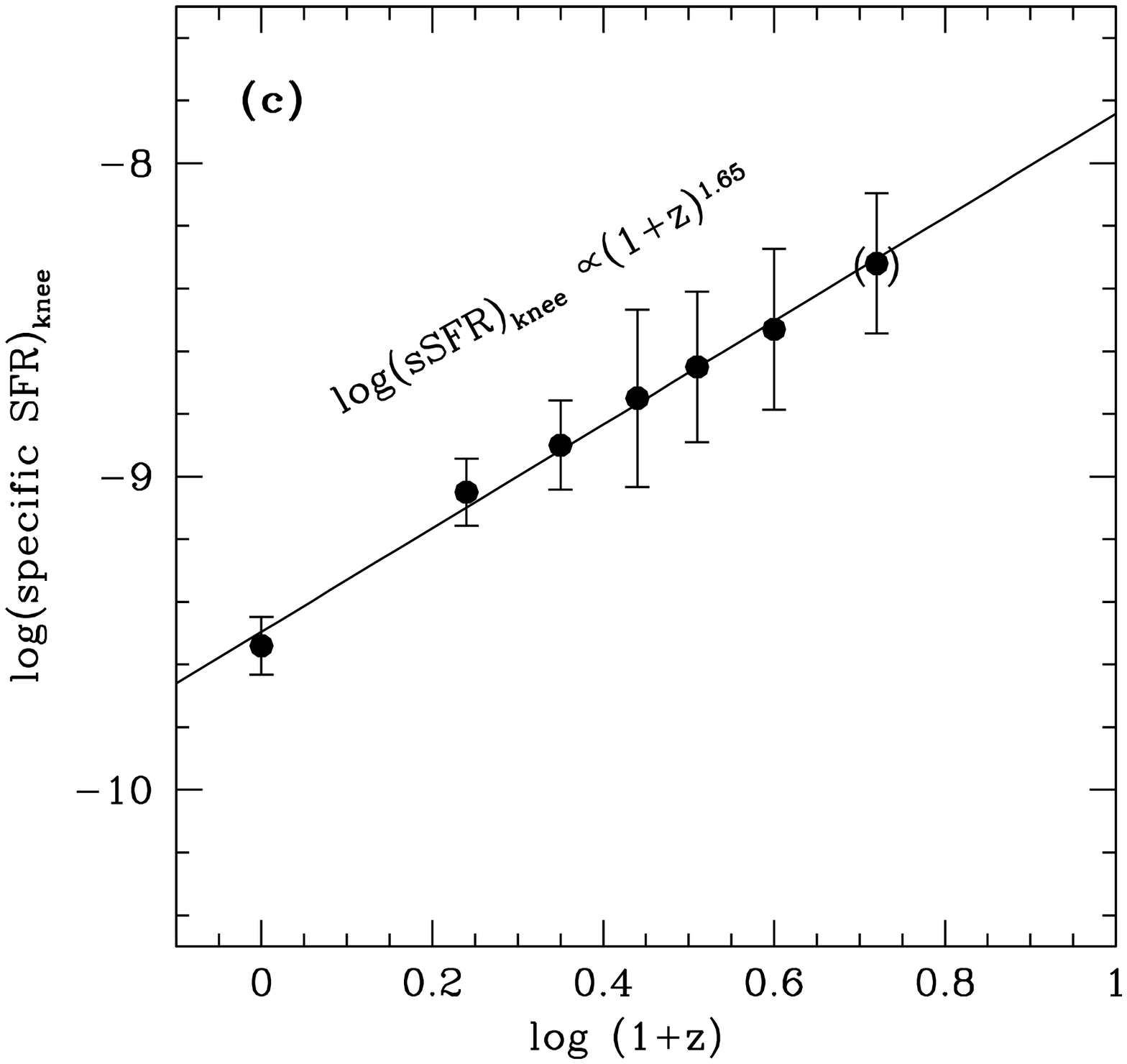}
\includegraphics[width=9cm, trim = 0cm 0cm 0cm 0cm]{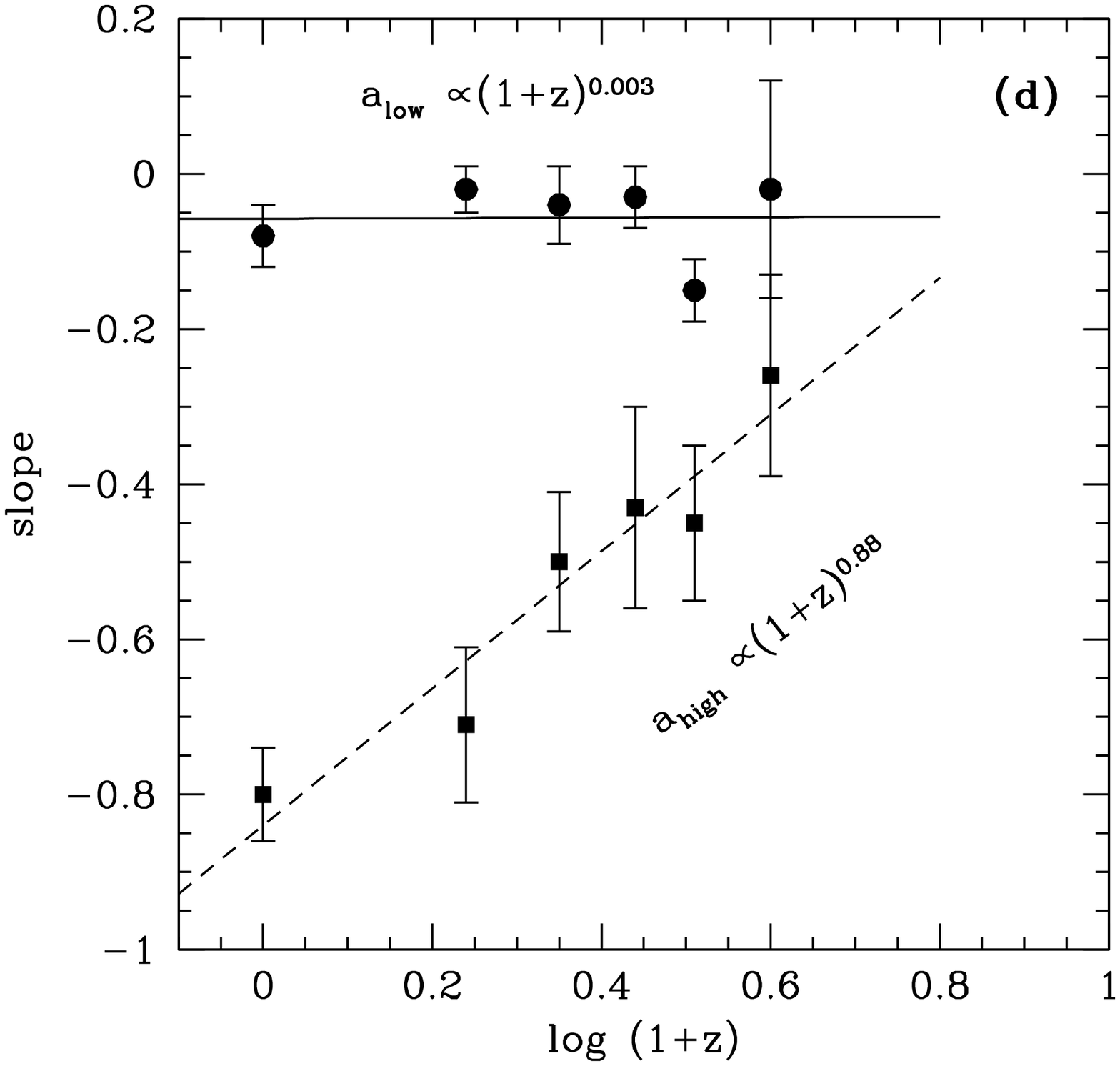}
\caption{(panel a): the specific star formation rate as a function of stellar mass in bins of redshift.
 For all redshift bins  the data are fitted with a broken power law with slope $a_{\rm low}$,
 holding below a critical mass ($M_{\rm knee}$), and  $a_{\rm high}$ holding above $M_{\rm knee}$ (see Table \ref{tab2}).
 The blue line connects the loci of $M_{\rm knee}$ for the various redshifts. 
(panel b): the position of the $M_{\rm knee}$ as a function of log(1+$z$).
(panel c): the specific star formation at $M_{\rm knee}$.
(panel d): the slope  below  and above $M_{\rm knee}$ ($a_{\rm low}$ and $a_{\rm high}$). 
The position of $M_{\rm knee}$ and the specific star formation rate at $M_{\rm knee}$ increase approximately as $(1+z)^2$, while
the mass quenching (given by $a_{\rm high}$) becomes less efficient with increasing redshift.
}
\label{ssfr}  
\end{figure*}

Figure \ref{ssfr} is derived from Figure \ref{sfr} after 
computing the sSFR at each redshift.  This figure
highlights that in most redshift bins (except for $z$=2.25) the specific star
formation rate is constant up to a characteristic stellar mass
($\ M_{\rm knee}$), beyond which it decreases steeply with increasing
stellar mass (Kauffmann et al. 2003).  In other words main-sequence star-forming galaxies 
above $ M_{\rm knee}$ have their sSFR suppressed compared to the lower mass
systems. Still, they remain classified as UVJ active galaxies, 
i.e., they are only partially quenched, and should not be confused with a passive population.

Similarly to the analysis by Whitaker et al. (2014), we fit to the sSFR versus mass relation a broken power law
of the form
\begin{equation}
\log {\rm sSFR} = a [\log (M_*/M_\odot)-\log(M_{\rm knee}/M_\odot)]+b,
\end{equation}
where $a = a_{\rm low}$ for $M_*< M_{\rm knee}$ and $a = a_{\rm high}$ for
$M_* \ge M_{\rm knee}$. In this equation, $b$ represents the sSFR at M$_{\rm knee}$.  
During the fit $a$, $b$, and $M_{\rm knee}$ are kept as free parameters, and the best-fit value 
is given in Table \ref{tab2}.
The resulting functions, which are plotted in Figure \ref{ssfr}(a), 
are found to be consistent with the results of Whitaker et al. (2014), 
even though we have kept $ M_{\rm knee}$ as a free parameter.  
Our approach allows for the study of the dependence of $M_{\rm knee}$ on redshift, 
which is found to be consistent with a scaling-relation $M_{\rm knee} \propto (1+z)^{2}$ 
as is shown in Figure~\ref{ssfr}(b)\footnote{An independent study (Lee et al. 2015) has recently been found that shows a similar trend}.  
This implies that any quenching mechanism 
at work within the main sequence becomes effective above some mass threshold, 
which decreases by more than a factor of 10 from $z$=3 to the present. 
We emphasize that, by construction, our analysis is insensitive to quenching mechanisms 
that would remove galaxies from the star-forming sequence altogether, while it is sensitive
to those mechanisms that perturb only in part the SFRs of main-sequence galaxies.
Figure \ref{ssfr}(c) shows how the sSFR evaluated at
$M_{\rm knee}$ scales with redshift, implying that the typical sSFR of the main sequence
depends on $z$ at the $1.65^{\rm th}$ power.  In turn, this implies a decrease by more than
one order of magnitude of the mean sSFR from $z$=4 to $z$=0 for normal
(unquenched) galaxies.  Figure \ref{ssfr}(d) shows the dependence on redshift of the
slope of the sSFR versus mass  relation below ($a_{\rm low}$) and above ($a_{\rm
  high}$ ) $M_{\rm knee}$. The parameter  $a_{\rm low}$ is independent of redshift, while
$a_{\rm high}$ increases as $(1+z)^{0.88}$: i.e., the main sequence of unquenched
galaxies exists at all redshifts, but the effects of quenching are less severe
with increasing redshift. This is in agreement with the findings of Whitaker et al. (2014).

\begin{table}[t]
\caption{Parameters of the fit for the function  $\log$ sSFR= $a (\log M_* - \log M_{\rm knee}) + b$}
\centering
\begin{tabular}{c c c c c c c c c}
\hline\hline
  $<z>$  &   $\log M_{\rm {knee}}$         & $a_{\rm low}$      & $a_{\rm high}$    &      b               \\
         &                        &                    &                   &                    \\
\hline   
  0.0  &  $   9.45\pm0.08  $ & $  -0.08\pm 0.04$ & $ -0.80 \pm    0.06$ & $-9.54 \pm      0.10  $ \\
  0.75 &  $  10.06\pm0.08  $ & $  -0.02\pm 0.04$ & $ -0.71 \pm    0.09$ & $-9.07 \pm      0.11  $ \\
  1.25 &  $  10.23\pm0.11  $ & $  -0.03\pm 0.05$ & $ -0.50 \pm    0.09$ & $-8.91 \pm      0.14  $ \\
  1.75 &  $  10.40\pm0.21  $ & $  -0.03\pm 0.04$ & $ -0.43 \pm    0.13$ & $-8.75 \pm      0.28  $ \\
  2.25 &  $  10.55\pm0.18  $ & $  -0.15\pm 0.04$ & $ -0.45 \pm    0.10$ & $-8.66 \pm      0.24  $ \\
  3.00 &  $  10.75\pm0.20  $ & $  -0.02\pm 0.14$ & $ -0.25 \pm    0.13$ & $-8.54 \pm      0.26  $ \\
\hline
\end{tabular}
\label{tab2}
\end{table}

\section{Strong bars and bulges as a function of $M_*$}
\label{bars}

In the previous section, we  determine  that galaxies above a 
redshift-dependent mass threshold are progressively more quenched.  It is necessary to
study in greater depth what physical mechanism might have caused such an
effect. 

We begin by taking a closer look at the morphology of the studied galaxies
below and above $M_{\rm knee}$, starting from the local sample.  
As discussed in the literature, it is quite challenging to produce 
a reliable morphological classification that distinguises disks
from bulges and, possibly, classical from pseudo-type bulges (Wilman et al. 2013).  
The task is even harder as recent evidence indicates that the two bulge categories 
can even occur simultaneously (Erwin et al.  2015). 

  With these caveats in mind, we focused on the detection of ``strong bars''
  (using the nomenclature of Nair \& Abraham 2010). We instead refrain from
  classifying ``weak'' and even ``intermediate'' bars as ``bars'', because we expect
  that they produce only minor perturbations to the disk, making them
  difficult to recognize.  The criteria used to visually identify strong bars
  include that the bar ellipticity must be larger than $\sim$ 0.4, but we did
  not impose any constraint on the galaxy maximum inclination. Of course bars
  are easier to detect in face-on systems, although the presence of X-shaped,
  boxy, or peanut shaped bulges helps detect bars even in highly inclined
  objects.  Secondary features such as rings near corotation, ansae, dust
  lanes, and inner Lindblad resonances (ILR, mostly too small to be detected on
  SDSS images) are not mandatory features, but of course - if present - they
  help in identifying bars.

  The visual classification of  strong bars was performed by seven authors  
  (GG, GC, MD, RF, MFo, MFu, GS)   who individually inspected and
  classified all  864 galaxies in our sample.

  The classification was based on
  $i$-band SDSS images, not to be biased by the sSFR versus color
  relation, nor by dust attenuation effects. 
  Following a template, the classifiers were called to distinguish
  i) barred, ii) unbarred galaxies hosting a bulge, and iii) disks without a bar or a
  bulge. Among class ii) we do not try to disentangle pseudobulges from classical bulges.

  Despite the aforementioned difficulties, the robustness of the resulting
classification is satisfactory overall: among the galaxies identified as hosts of a
strong bar, agreement between more than four classifiers was reached in 92\% of the cases; 
the level of agreement drops to 77\% for bulges and to 85\% for disks without bars or bulges. 
These percentages suggest that the main difficulty lies in the identification of bulges,
reflecting the ambiguity in detecting the presence of bulges in face-on or poorly resolved disk 
galaxies when the color information is disregarded (see also Drory \& Fisher 2007).
  
\begin{figure*}[!t]
\centering
\includegraphics[width=18.3cm, trim = 0cm 0cm 0cm 0cm]{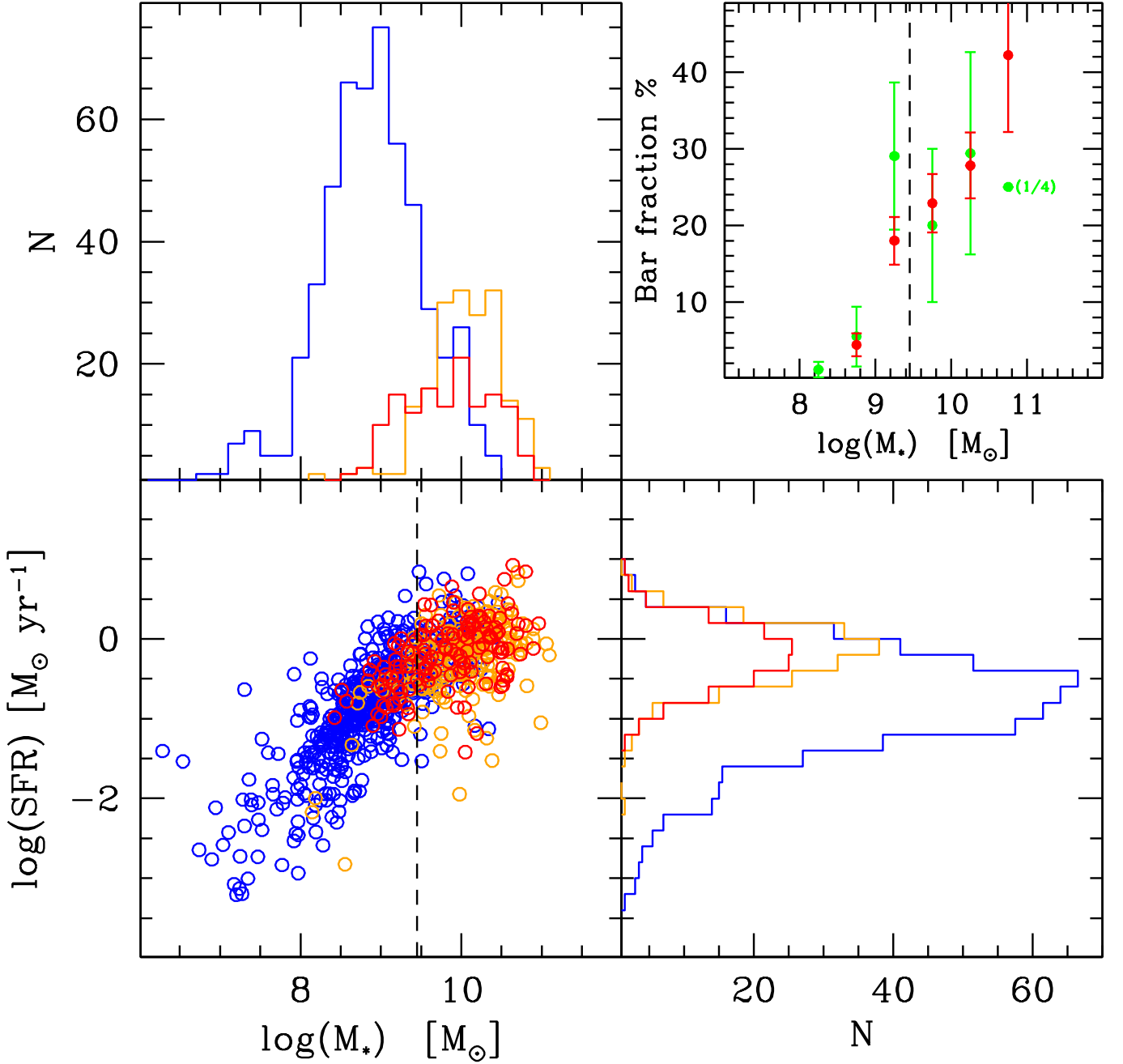}
 \caption{The SFR versus stellar mass for the local sample in Figure 1, but 
  with symbols corresponding to the visual morphology:
  disks without bulge or bar  (blue),  barred disks (red), and unbarred disks with a bulge (orange). 
  The vertical dashed line indicates  $M_{\rm knee}$ at $z$=0.
  Within the same categories, distributions in bins of stellar mass and SFR are given in the top and 
  right histograms. 
   The top right panel gives the fraction of visually classified strong bars as a function of stellar mass,
  given separately for the whole sample (red), and for the local sample (within the distance of 40 Mpc, green).
  Owing to the small sampled volume, the third subsample lacks  statistical weight at high mass (one barred galaxy over 4
  objects in the highest mass bin), while the point at the low-mass end has the highest statistical significance (one barred
  galaxy over 85 targets). 
  }
 \label{barre}  
\end{figure*}

  The result of this morphology classification is shown in Figure
 \ref{barre}, which presents again the SFR versus stellar mass relation that is
 now color coded according to the morphological classification. The value of $M_{\rm knee}$ is 
 indicated by the vertical dashed line. 
 Histograms are
 also provided to highlight the relative frequency of each class in bins of
 mass and SFR. 
 
 Below $M_* = \rm 10^{9.45}$ M$_\odot$, i.e., $M_{\rm knee}$ at $z$=0, the 
 frequency of disks without bulge or bar, barred disks, and unbarred disks with a bulge 
 is 87\%, 8\%, and 5\% respectively.  
 Above $M_* = \rm 10^{9.45}$ M$_\odot$, instead, these frequencies
 become 28\%, 27\%, and  45\%. Our analysis reveals that the vast majority of low-mass galaxies 
 are disks without bulge or bar, while more than half of the high-mass galaxies host either a bulge or a strong bar.
 
 In the top right panel, we plot  the occupation
 fraction   of visually classified strong bars of the whole sample as a function of stellar mass.
 This is in agreement with previous studies 
 highlighting that the likelihood of having a bar in disk galaxies increases with increasing stellar mass.
 Skibba et al. (2011), Wang et al. (2012), and Masters et al. (2012)  consistently find that the  strong-bar 
 fraction increases from 10\% to 40\% with
 increasing stellar mass from $M_*=10^9$ to $M_*=10^{11}$ M$_\odot$. Consistent conclusions are indirectly reached by
 Marinova et al. (2009) (who study the dependence on V luminosity).
 %Sheth et al. 2012) (decrescita bar fraction con redshift esclus alle basse masse).  
 Moreover, focusing on the nearby Virgo cluster and using the early bar classifications by de Vaucouleurs et al. (1991, RC3)  and by Binggeli et al. (1985)
 from high-quality photographic plates,
 we find that in the Local and Coma superclusters the fraction of barred galaxies is lower than 20\%
 below $M_* = \rm 10^{9.5}$ M$_\odot$ and rises to 30-40\% at high masses. Again, this is
 consistent with all results listed above.  \\

 On the other hand, two other results contradict this trend.
 Barazza et al (2008) and Nair \& Abraham (2010) found bar fractions on  the order of 30-40\% above
 $M_* = \rm 10^{10}$ M$_\odot$, consistently with all cited works, but their strong-bar fraction 
 increases  with decreasing mass, reaching 40-50\% around $M_* = \rm 10^{9.5}$ M$_\odot$.
 We note, first of all, that these authors
 did not include dwarf irregulars in their study, but in fact these galaxies represent the majority in our sample among low-mass galaxies.
 We deliberately included them as they are gas-rich, star-forming main-sequence objects obeying  the Tully-Fisher relation
 % Moreover at low-mass their bar population is dominated by weak bars which, instead we disregard as they hardly produce major perturbations.
 and this could explain the large discrepancies between our work and theirs.
  We stress that by selection, galaxies shown in Figure \ref{barre} include dIrrs but not dEs. This makes a direct comparison
 to other studies of the frequency of strong bars as a function of mass, such as Nair \& Abraham (2010) and Barazza et al  (2008), more difficult.

%\subsection{Caveats}
To further prove our point we checked against possible biases that could in principle artificially reduce the frequency of bars especially at low mass. 
These are traceable to the following cases:
 i) obscuration by dust;  ii) galaxy inclination; and iii) limited spatial resolution (hampering the detection of bars in small galaxies 
 and in gas-rich galaxies with patchy star formation).

 Case i) It has been shown by several groups that 60\%                                                                                                                                                           
 of bright disk galaxies are barred in the near-infrared (Eskridge et al. 2000; Laurikainen et al. 2004; Menendez-Delmestre et al. 2007; Marinova \& Jogee 2007),                                       
 while only ~45\% appear barred in the optical (Eskridge et al. 2000; Reese et al. 2007; Marinova \& Jogee 2007), presumably due to dust obscuration.                                                           
 Our lack of bars at low stellar masses in Figure 4 could result from this dust bias if more obscuration associated with a larger dust fraction                                                                
 occurs at lower mass. However, first our bar selection band (SDSS $i$ $\lambda=7600 \AA$) is  closer to near-infrared than other bluer optical bands;                                     
 second, we used a representative sample of the local universe (the HRS sample of Boselli et al. 2010) to compute the                                                                               
 extinction coefficient as a function of stellar mass and (unsurprisingly, given the mass-metallicity relation) we found that it decreases with decreasing mass,                                              
 such that for mass below $M_*=10^{9}$ M$_\odot$  $A(i)<0.2$ mag (Boselli et al. 2015), ruling out a strong obscuration at low mass, and consequently a possible                                         
 bias in our ability to find bars at the low-mass end of the distribution.                                                                                                 
                                                                                                                                                                                                            
 Case ii) We found no significant                                                                                                                                                                                
     bias related to the galaxy inclination and/or to the relative PA between the                                                                                                                         
     bar and the inclined galaxy. Qualitatively, while bars could be missed in very inclined systems, 
     there is no reason to expect a higher incidence of bars in inclined systems at lower masses. More quantitatively, we computed 
     the fraction                                                                                                                                   
     of strong bars as a function of stellar mass in a subsample of face-on  ($i < 45^o$) galaxies,                                                                                             
     and checked that the bar fraction  remains unchanged as a function of mass. Actually, by                                                                                                          
     adopting the same mass bins  as used in Fig. \ref{barre} (from 8.5 to 11.0 in steps of 0.5 $\log(M_\star$)),                                                                                              
     the bar fractions become 4\%, 11\%, 25\%, 26\%, and  35\%, which are consistent                                                                                                                             
     with the results obtained when analyzing the  whole sample (4\%, 17\%, 23\%, 28\%, 42\%), confirming that our results are not affected                                                                   
     by any inclination bias\footnote{Our sample is also not biased by the relative PA                                                                                                                          
      as it does not affect  face-on galaxies in any way.}.                                                                                                  
                                                                                                                                                                                                            
 Case iii) Our sample is limited to $z<0.03$, but in the high-mass range it is dominated by                                                                                                                        
   objects with $z \sim 0.02$ (Coma Supercluster), while at low mass Local Supercluster galaxies dominate.                                                                                             
   The spatial resolution offered by SDSS images ($\sim 1.4$ arcsec) corresponds to 0.7 kpc at the distance of Coma.                                                                                         
   As discussed by Barazza et al. (2008), the typical scale of strong bars at high mass (above $5 \times 10^9$ M$_\odot$) is 2 kpc,                                                                               
   which does not hamper the bar detection at high mass.                                                                                                                                                       
  At low mass (below $10^9$ M$_\odot$) our sample is instead dominated by dIrr in the Local Supercluster                                                                                              
   where the SDSS resolution element becomes $\sim 110$ pc, i.e., sufficient to resolve bars whose size is                                                                                                      
   10\% of their optical diameter (Erwin et al 2005), which is typically 2.5 kpc.                                                                                                                        
  In order to test the robustness of the determination of the bar fraction with respect to the spatial resolution,                                                                                             
  we split the sample in two distance bins: within 40 Mpc (i.e., dominated by the Local Supercluster), and one between                                                                                         
  40 and 100 Mpc (i.e., dominated by the Coma Supercluster). In Figure \ref{barre} we plot  the bar fraction in the nearby subsample                                                                           
  (green dots) separately from the total (red dots).

  %The left panel of Fig. \ref{unbarred} reports the gallery of the 9 dwarfs ($M_* < \rm 10^{9}$ M$_\odot$) galaxies hosting a strong bar. 
  %The right panel of Fig. \ref{unbarred} reports a gallery of 9 randomly selected unbarred galaxies in the same mass range out of a total of 312 unbarred objects. 

  At low mass ($M_* < \rm 10^{9}$ M$_\odot$), nine dwarf  galaxies host a strong bar, while 
  in the same mass range 312 objects are classified as unbarred. 
  None of them appears to have a missed bar; however, 
  among these 312 candidates, 7 galaxies received at least one bar-vote from one of the classifiers. 
  This would bring the bar fraction below $10^{9}$ M$_\odot$ to at most 5\%, 
  significantly below the frequencies measured by Barazza et al (2008) and Nair \& Abraham (2010).

We finally check the dependence of the bar fraction on color. Given the known color-mass relation, e.g., more massive galaxies exhibit redder colors, it is not surprising that
Skibba et al. (2011) and Masters et al. (2012) find that the bar fraction increases from 10 to 40\% from blue to red, while
Barazza et al (2008) and Nair \& Abraham (2010) do not find such an effect.
In our sample the bar fraction is 13\%, 16\%, 25\%, and 21\% with $g-i$  increasing from 0.25 to 1.25 in steps of 0.25.
Above $M_{\rm knee}$, bars are undoubtedly associated                                                    
with red regions, as vividly demonstrated by Figure \ref{5921}, where a picture of the barred  
galaxy NGC 5921 is shown. Within the bar extent (red circle) the                                         
color index is as red as the color of an early-type galaxy (ETG), while it is as blue as a          
typical massive LTG outside this radius.  This color pattern is the same as                            
 the other massive barred galaxies in our local sample, as shown by the                              
red lines in the right panel of Figure \ref{5921}. These lines correspond to                               
the median color profiles of barred galaxies with mass above $M_{\rm knee}$                              
and different inclination cuts. Profiles have been normalized to the bar length.                  
Despite the projection effects that smear  the sharp color gradient seen in NGC 5921                         
a change in the color profile is still visible near  the bar edge because, even in face-on galaxies, the zone containing the bar often has                
a higher ellipticity and a position angle                                                                
that is different from that of the galaxy as a whole, which is used to compute the color profile.

\section{Bar-driven star formation quenching}
\label{theory}

%Our observational study shows that: $(i)$ the low sSFR of massive galaxies is
%due to the presence of a central (kpc-scale) quiescent/quenched region, often
%still surrounded by significant star formation; $(ii)$ the redder central
%region is associated to a red disc like structure in which, about half of the
%times, a clear bar is present. In the remaining $\sim 50 \%$ of the sample the
%images are consistent with a pseudobulge structure hosted in the galactic
%nucleus; $(iii)$ the mass at which the sSFR starts to decline depends on the
%redshift of the galaxy sub-sample we consider.

In this section we propose a simple model in which a forming or existing bar
removes in  few dynamical times most of the gas from the central region of the
galaxy (i.e., within the bar corotational radius). As a consequence, after
a short transient nuclear starburst, the inner region of the galaxy stops
forming stars, and grows redder with time  (see also Cheung et al. 2013). This model provides a
simple and natural explanation of our observational evidence presented so
far. We note, however,  that our model  applies only to isolated disk galaxies. 
Dynamically hot stellar systems,  elliptical for example,  would not form bars, and other environmental processes
are known to act on galaxies in clusters (Boselli \& Gavazzi 2006).

At first, in Section~\ref{simulations}, we consider a single bar-unstable
galaxy, and, through the comparison with a numerical simulation, we show that
the main features of massive disk galaxies observed are nicely reproduced even
with the most simplifying assumptions.  Then, in Section~\ref{dynamics}, we
make use of simple analytical considerations to demonstrate that the proposed
model of bar-driven quenching reproduces the observed dependences of the sSFR
on the galaxy masses and redshifts.

\subsection{Comparison with hydrodynamical simulations}
\label{simulations}

\begin{figure*}[!t]
\centering \includegraphics[width=6.9cm, trim = 0cm 0cm 0cm 0cm]{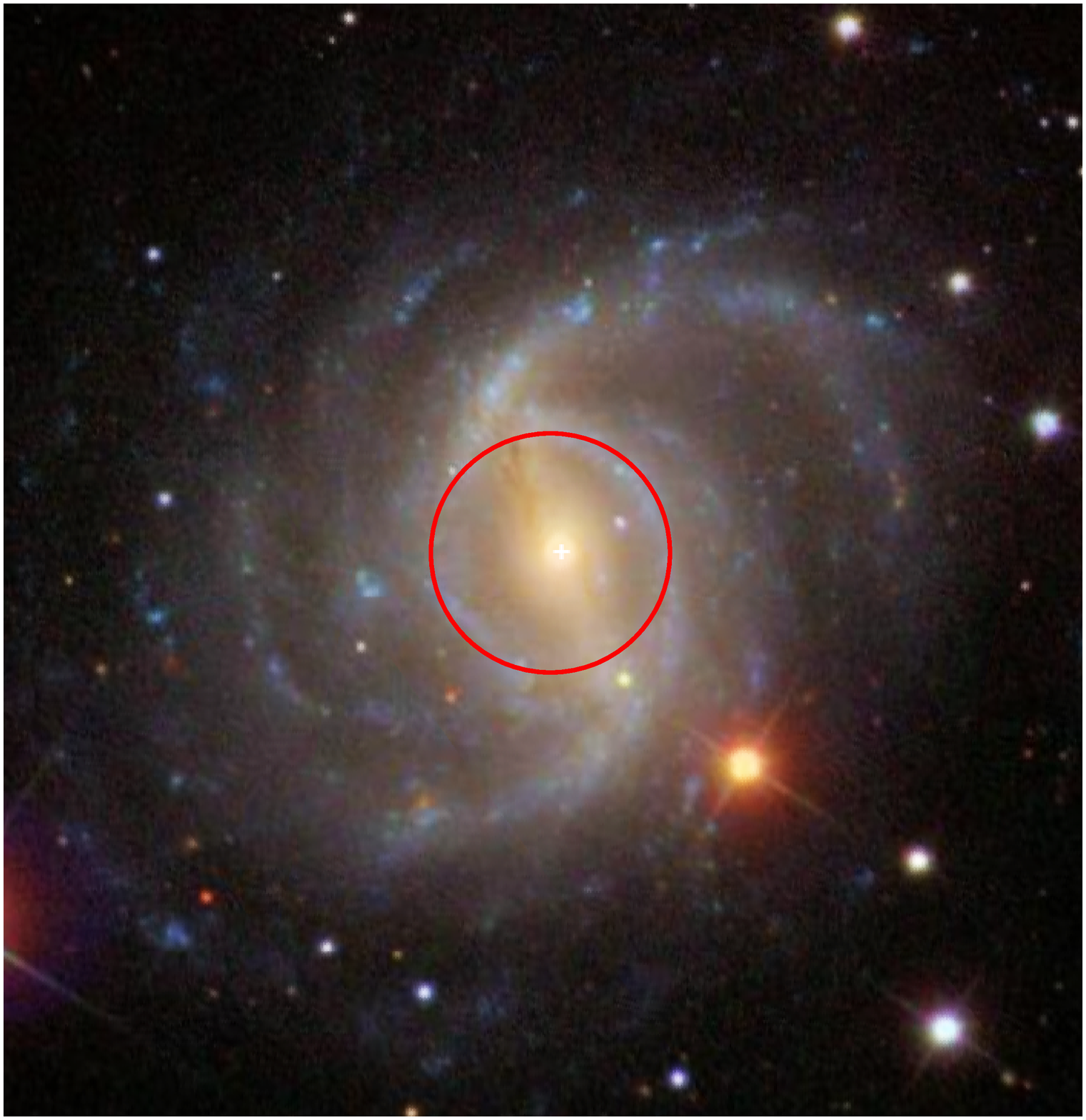}
\hspace{5pt} \centering \includegraphics[width=9.5cm, trim = 0cm 0cm 0cm 0cm]{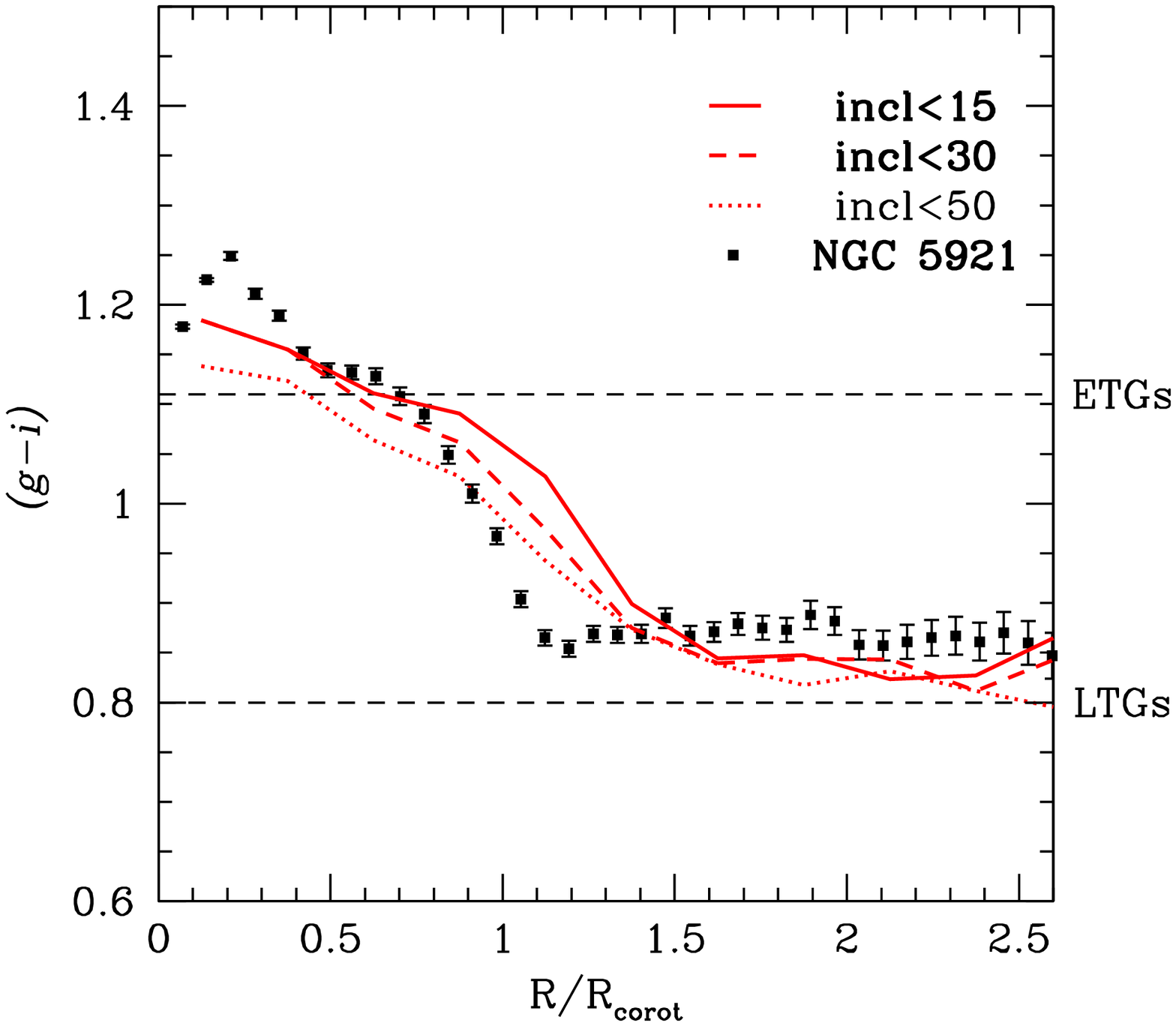}
 \caption{RGB image (SDSS) of the barred galaxy NGC 5921 (left panel). 
 The bar extent is marked in red. Its $g-i$ color profile along the major axis in units of 
 corotation radius (dots) is superposed to the median color profiles
 of barred galaxies with mass above $ M_{\rm knee}$ and different inclinations ($<$15 deg, 30 deg, 50 deg).
 The two dashed horizontal lines mark the color of typical LTGs and ETGs above $M_{\rm knee}$.
 }
 \label{5921}  
\end{figure*}

\begin{figure*}[!t]
\centering \includegraphics[width=18.3cm, trim = 0cm 0cm 0cm 0cm]{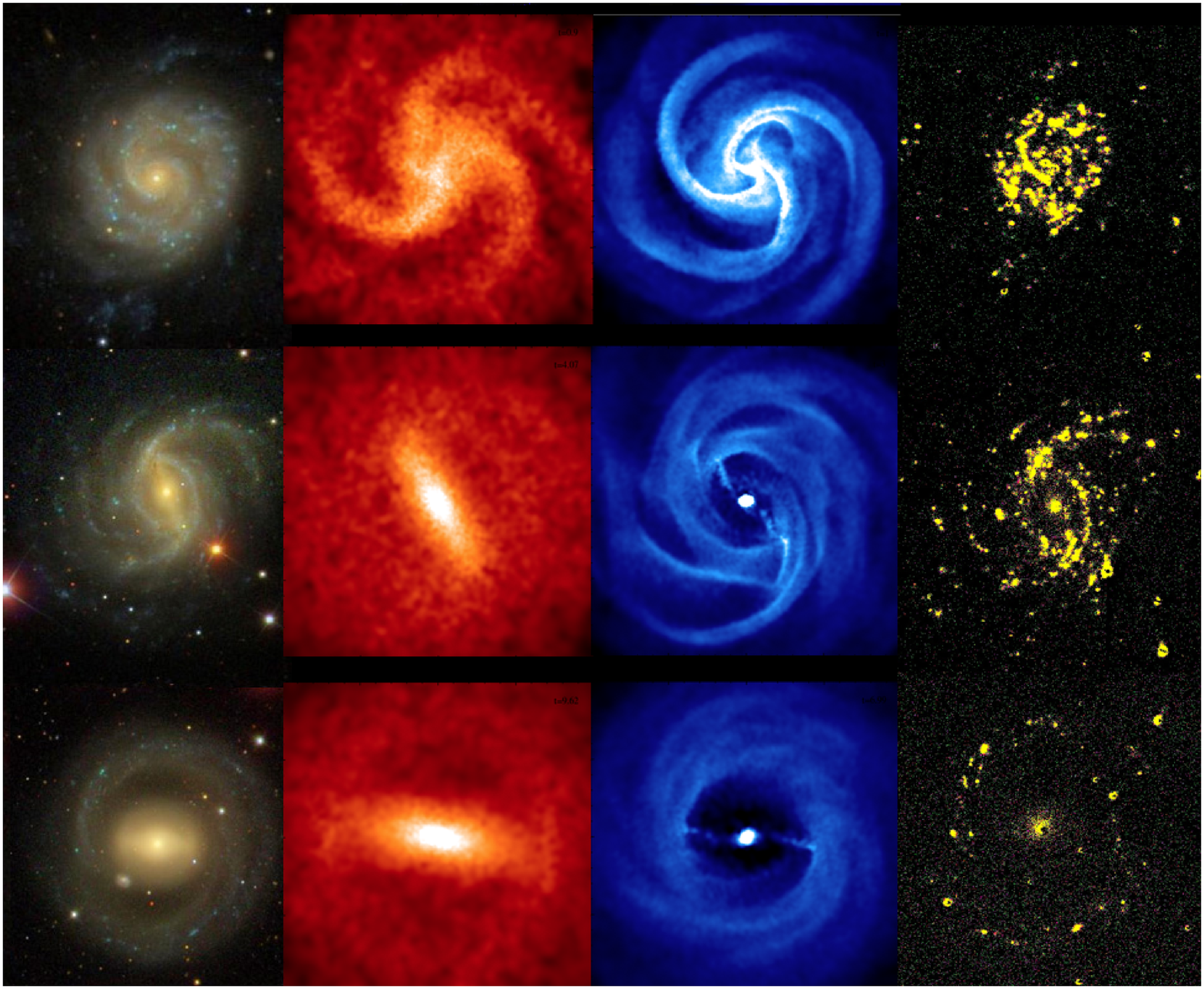}
 \caption{RGB images (SDSS) of three galaxies (left column) with increasing
   stellar mass (NGC 3596, NGC 5921, NGC 5701 from top to bottom) showing a
   regular spiral galaxy (top), a well-developed bar (middle), a barred ring
   (bottom). Face-on views of simulations of stars (second column) and gas
   densities (third column) from $t$=1 to $t$=4 Gyr showing a spiral disk galaxy (top)
   that becomes bar unstable (middle). Inside the corotation radius, 
   the gas is conveyed towards the center and quickly consumed 
   (except for little left along the bar). 
   Outside the corotation radius, the gas is unperturbed and feeds
   peripheral star formation. At the latest time step ($t$=9 Gyr) the galaxy
   fully develops its bar and the central region is completely evacuated of
   gas. A ring of gas is left outside and feeds the star formation, as confirmed
   by the rightmost column showing our H$\alpha$ images.  When the bar is
   well developed (the two bottom panels) the star formation is suppressed
   inside the bar corotation radius, but it is ongoing in the
   outer parts. Even centrally quenched galaxies host some  emission on a nuclear scale 
   (rightmost column). However, we caution  that this emission is dominated by 
   [NII] with respect to H$\alpha$ (common in LINERs) 
   and so does not indicate solely ongoing star formation.}
 \label{simul}  
\end{figure*}

   As a test-bed for the study of the effects of a strong bar on the gas on
  galactic scales we analyze one of the N-body/hydrodynamic simulations of
  isolated disk galaxies discussed in Fanali et al. (in prep.). In this run no
  star formation prescription or any kind of star formation/AGN related
  feedback has been implemented in order to allow for a clear identification
  of a dynamical quenching effect of the bar, if present. Reassuringly, despite the
  simple numerical techniques adopted in our calculation, 
  the results discussed here are in line with the findings of other authors, as we
  will detail in the following.

The initial conditions of the run are equal to those described in Mayer \&
Wadsley (2004, model Lmd2c12), in order to reproduce an initially bulgeless
bar-unstable galaxy.  The stellar component of the galactic disk follows an
exponential profile
\begin{equation}
\rho_* (R,z) = \frac{M_*}{4 \pi R_*^2 z_*} \exp(-R/R_*) {\rm sech}^2(z/z_*),
\end{equation}
where the radial and vertical scale lengths are $R_*=3$ kpc and $z_*=0.3$ kpc,
respectively, and $M_*= 1.4 \cdot 10^{10}$M$_\odot$ is the total stellar mass.
The galactic disk has an additional gas component of mass $M_{\rm gas}=f_{\rm
 gas} \times M_{*}$, with a gas fraction $f_{\rm gas} = 0.05$. The gas
follows the same surface density profile of the stars, and it is assumed to
have a  homogeneous temperature profile, with $T_{\rm gas}=10^4$ K. The gas
evolves isothermally during the system evolution.  We will see that our
simulation reproduces all the key features of massive disk galaxies that we need to
test our model, even under such simple assumptions about the gas thermodynamics.

The composite stellar-gaseous disk is embedded in a larger scale dark matter
halo, following a density profile
\begin{equation}
\rho (R)=\rho_H \frac{\delta_c}{(R/R_s)(1+R/R_s)^2},
\end{equation}
where $R_s=10$ kpc, $\rho_H$ is the critical density of the Universe today and
$\delta_c =(200/3) \times\{c^3/[ \ln (1+c)-c/(1+c)]\}$ depends only on the
concentration parameter $c$, set equal to 12 for this galaxy (Navarro et
al. 1995).  

   For each component (halo and disk) the particle positions are generated
  through a direct Monte Carlo sampling of the density profiles. Because of the
  complexity of the system, we do not generate the particle velocities by
  directly solving  the collisionless Boltzmann equation. We instead enforce an
  approximate dynamical equilibrium for the system, following Hernquist (1993,
  H93 hereafter) and Springel et al. (2005). In detail, we make use of the
  Jeans equation to compute the first and second
  moments of the velocity field as a function of the position, i.e., the bulk motion of the particles and the
  components of their local velocity dispersion. In the simpler halo case we
  assume an isotropic velocity field (i.e., no net rotation) and that all the
  components of the velocity dispersion tensor are equal.  The three
  components of the velocity dispersion tensor as well as the rotational bulk
  velocity of the disk particles are obtained following the numerical
  procedure described in Section~2.2.3 in H93. The velocity components of each
  particle, then, are sampled through a Monte Carlo procedure, assuming that the
  local distribution function is Gaussian, in good agreement with the
  observational constraints (see the discussion in H93).

We sample the stellar disk with $9.5\times 10^5$ particles, the gaseous disk
with $5\times 10^4$ particles, and the halo with $10^6$ particles. We ensure
that the particles in the disk all have  the same mass, preventing any spurious
relaxation and mass segregation. The softening length that sets the spatial
resolution of the gravitational interaction in the run of each particle is 15
pc. The system is evolved using the smoothed particle hydrodynamics (SPH) code
Gadget-2 (Springel 2005).

Three snapshots of the stellar and gas surface densities at different times
are shown in the two central columns of Figure~\ref{simul}, together with
three images of real galaxies taken from our sample for comparison (left
columns). The simulated galaxy has a first evolutionary phase ($t<1.5$ Gyr)
during which it develops mainly spiral features (top panels in
figure~\ref{simul}).  At $t\sim 1.5$ Gyr a stellar bar forms, and during its
growth and evolution it triggers strong gas inflows toward the galaxy
center. Already at $t\sim 4$ Gyr (second row in in Figure~\ref{simul})
  most of the gas in the central 4.5 kpc has been forced into the galactic
  nucleus, in accordance with previous observational (e.g., Sakamoto et
  al. 1999, Jogee et al. 2005, Sheth et al. 2005) as well as analytical and
  numerical studies (e.g., Sanders \& Huntley 1976, Shlosman, Frank \& Begelman
  1989, Athanassoula 1992; Berentzen et al. 1998; Regan \& Teuben 2004, Kim et
  al. 2012, Cole et al. 2014).

Although our simulation does not include any prescription for star formation,
the extreme gas densities in the nucleus and  its short dynamical time
ensures that most of the gas mass is doomed to convert into stars in a burst
of nuclear star formation (e.g., Krumholz et al. 2009, Krumholz \& McKee 2005,
Daddi et al. 2010, Genzel et al. 2010), likely resulting in the
formation of a pseudobulge.\footnote{ The higher occurrence of strong
  episodes of nuclear star formation in barred galaxies has been extensively
  observed, see, e.g., Ho et al. (1997), Martinet \& Friedli
  (1997), Hunt \& Malkan (1999), Laurikainen et al. (2004), Jogee et
  al. (2005).}  After the short transient starburst event, the gas density
(and, consequently, any expected star formation rate) drops. After $9$ Gyr
the stellar bar has swept the quasi-totality of the gas in the central 4.5 kpc
(bottom row in in Figure~\ref{simul}), and our simulation nicely reproduces the
properties of a centrally quenched galaxy as NGC 5701, but retains an
evident external spiral structure (bottom left panel in Figure~\ref{simul}).
Streams of low-density gas falling along the edges of the bar are still
visible, both in the simulation and in the observations, where they are
traced by dust filaments. The H$\alpha$ images shown in rightmost column of
Figure~\ref{simul} show (from top to bottom) that normal star formation is
taking place in the disk of the relatively lower mass NGC 3596, while when the
bar fully develops (NGC 5921 and NGC 5701) the star formation activity is null
inside the bar corotation radius, but remains conspicuous outside it.  Some
emission remains observable in the nuclear regions, hosting a star-forming
cluster (in NGC 3596) or showing [NII] over H$\alpha$ ratios suggestive of
low ionization nuclear emission-line regions (LINERs, as in NGC 5921 and NGC 5701), 
as the nuclear spectra of these three galaxies indicate (Gavazzi et al. 2013c).

\begin{figure}[!t]
\centering
\includegraphics[width=9cm, trim = 0cm 0cm 0cm 0cm]{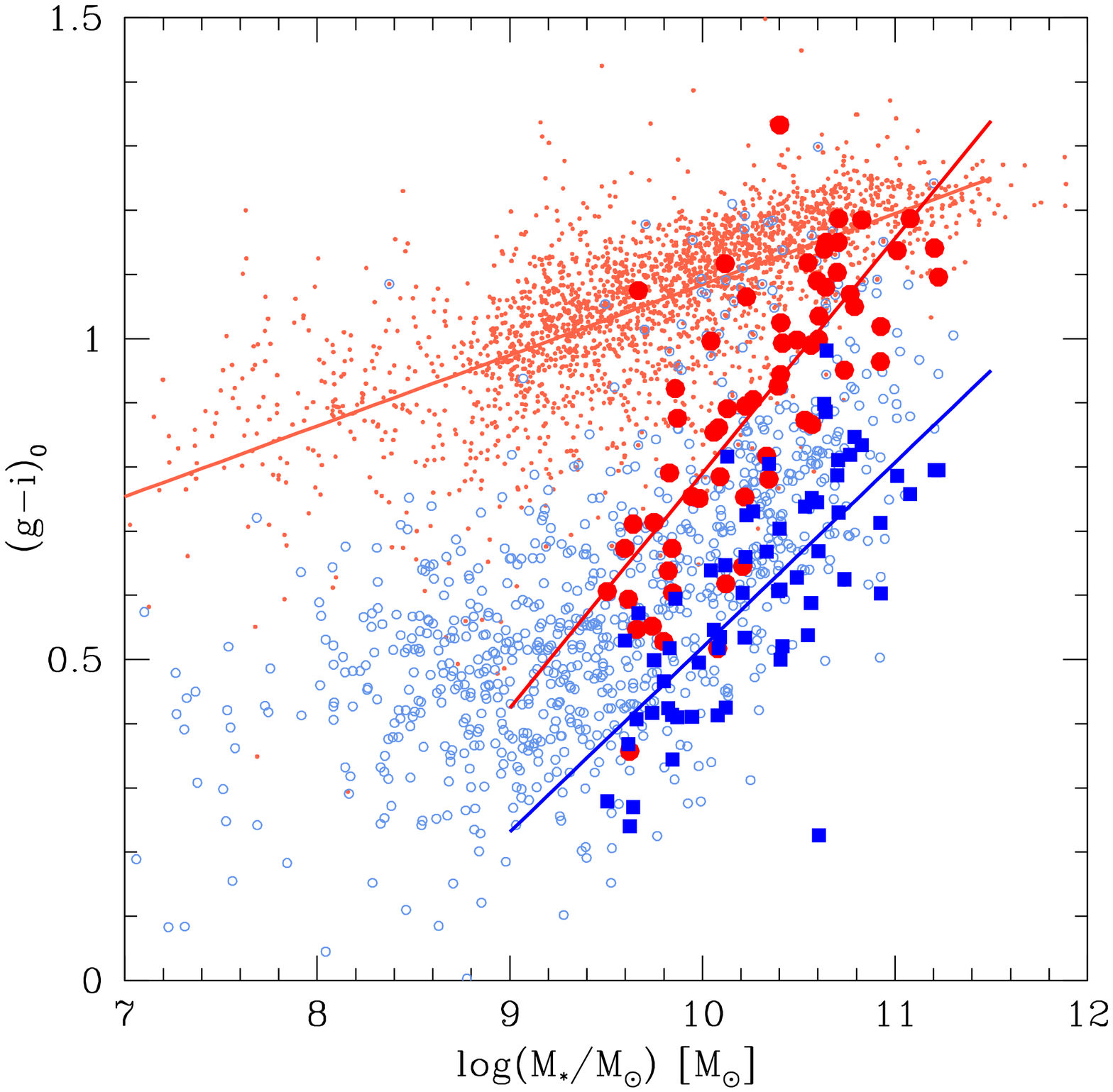}
 \caption{Color-mass diagram obtained with the $g-i$ color (corrected for
     extinction in the Milky Way and for internal extinction).  ETGs in the Coma and Local Superclusters (small red
     points) are also shown for comparison. The LTGs are subdivided into galaxies with disks without a bulge or a bar 
     (including
     dIrr and other blue dwarfs, light blues symbols) and galaxies that we classified as containing a strong bar. 
     The colors of the latter are separately displayed
     as large red symbols within the corotation radius and with large blue
     squares outside the corotation radius. Fits to the colors of the inner
     and outer regions of barred galaxies (as well as for ETGs) are shown
     in the figure.}
 \label{colmag}  
\end{figure}

The comparison between our simulation and observations have been
  performed for more than three galaxies. We note that Figure~\ref{5921}
  already demonstrates that the central regions of barred galaxies are, on
  average, quenched with respect to the corresponding outer parts.  To further
  support this scenario with our observational data of nearby galaxies, we
  present in Figure~\ref{colmag} the color-mass diagram dividing the inner
  parts (within the bar extent) of barred galaxies from
  their outer parts. The $g-i$ colors,
  taken as a proxy for sSFR, have been corrected
  for Milky Way and internal extinction as in Gavazzi et al. (2013b). Non-barred LTGs in our sample 
   and ETGs in the Coma and Local superclusters
  have been plotted for comparison. Figure~\ref{colmag} clearly demonstrates
  the significant central quenching caused by the bars in massive galaxies. We
  also note that the bar-driven gas removal cannot be the only quenching
  mechanism in place, as even the exteriors of massive barred galaxies are
  redder than lower mass counterparts. Our selection criteria, however, allow us to
  exclude a possible environmental nature of the additional quenching
  process.
We note that the triggering of a strong gas inflow like the one observed in
our simulation is ubiquitously observed in many investigations, regardless of
the particular type of code used (2D versus 3D, Eulerian grid based codes
versus SPH, see Sellwood 2014 for a thorough discussion).  As a final
  word of caution, we highlight that the main shortcut of our simulation is
  the lack of any feedback associated with star formation or to the possible onset
  of an AGN. This allowed us to firmly identify an independent -- purely
  dynamical -- bar driven quenching process.  Stellar feedback could, however,
  eject a significant amount of the gas driven into the galaxy central regions
  by the bar. Because of the small angular momentum of such outflows, the gas
  would not re-enrich the quenched kpc size region. The ejected gas is instead
  expected to fall back towards the nucleus, leading to multiple episodes of
  intense and fast nuclear star formation (see the results of the high-resolution simulation including stellar feedback discussed in Emsellem et
  al. 2015).

\subsection{Dynamical model for the sSFR-quenching cosmic evolution}
\label{dynamics}

In the previous section we discuss how the presence of a bar results in
the removal of  gas in the central region of a galaxy, explaining its red
colour within the corotation radius and the lower sSFR of the whole galaxy. 
Figure~\ref{ssfr}, however, clearly indicates that the bar-driven quenching is not
effective in  low-mass galaxies. More specifically, the observational data are
indeed consistent with strong bars forming only in massive spiral galaxies, with $M_*
> M_{\rm knee} \sim 10^{9.5}$ M$_\odot$ at  $z=0$, where $M_{\rm knee}$
is an increasing function of redshift.
We stress that the trend of $M_{\rm knee}$ with redshift fits with the results
of studies on the strong bar frequencies in large observational samples. As an
example, Sheth et al. (2008) analyzed the COSMOS 2 deg$^2$ field finding a
decreasing bar fraction moving toward higher redshifts.
They also comment on the fact that the strong bar fraction of low-mass $M_* \lsim 10^{10.5}$
M$_{\odot}$ spirals declines significantly with redshift beyond $z$=0.3, while
it remains roughly constant out to $z$ $\sim$ 0.84 in more massive, luminous
spirals.  This is consistent with the seminal results of Jogee et al (2004), obtained
analyzing galaxies out to $z \sim 1$ observed with the Hubble Space Telescope Advanced Camera for Surveys (ACS). 
Later on, similar results were discussed in Cameron et al. (2010) who reported that the strong bar fraction for massive systems 
($M_* > 10 ^{11}$ M$_\odot$) does not change between $z = 0.2$ and z$= 0.6$, while it falls for lower mass systems.
More recently, Melvin et al. (2014), on the  basis
of visual classifications provided by citizen scientists via the Galaxy Zoo
Hubble project, also find that the overall  strong bar fraction decreases 
from 22 $\pm 5$\% at $z$=0.4 to 11$\pm 2$\% at $z$=1.0. In addition, they confirmed
that this decrease in the bar fraction is most prominent at low stellar masses.

A simple model that explains the existence of a redshift-dependent 
$M_{\rm knee}$ above which the sSFR declines can be built on the observational 
evidence that the dynamical state of galactic disks depends on their masses and redshifts
(e.g., Sesana et al. 2014 and references therein). We start noticing that
dynamically hot galactic disks are stable against bar formation, while colder disks can 
form bars in few dynamical times (e.g., Athanassoula \& Sellwood 1986). 
We define hot disks those with Toomre parameter 
\begin{equation}
Q \sim \frac{\sigma_* \Omega}{G \Sigma_*} \gsim 1,
\label{eq:q}
\end{equation}
where $\sigma_*$ is the 
stellar velocity dispersion, $\Omega$  is 
the angular velocity and $\Sigma_*$ is the stellar surface density.
The Toomre parameter can be rewritten as
\begin{equation}
Q \sim \left(\frac{v_{\rm rot}}{\sigma_*}\right)^{-1}\,\left(\frac{v_{\rm
    rot}^2 r}{G M_*} \right)\sim \left(\frac{v_{\rm rot}}{\sigma_*}\right)^{-1}, 
\label{eq:q2}
\end{equation}
where $v_{\rm rot}$ is the rotational velocity of the disk, $r$ is a proxy for
the disk extension, and $(v_{\rm rot}^2 r)/(G M_*) \sim 1$ because of the
virial equilibrium of the rotating stellar disk. Using Eq.~\ref{eq:q2} we
can translate the critical value of the Toomre parameter $Q_{\rm crit}$,
distinguishing between bar stable and unstable systems, into a critical value
of the $v_{\rm rot}/\sigma_*$ ratio. 

A growing number of studies (e.g., F\"orster Schreiber et al. 2009; Law et
al. 2009; Gnerucci et al. 2011; Kassin et al. 2012; Wisnioski et al. 2011;
Epinat et al.  2012; Swinbank et al. 2012; Newman et al. 2013; Wisnioski et
al. 2015) finds that the $v_{\rm rot}/\sigma_{\rm gas}$ in disk galaxies
increases as a function of the galaxy mass $M_*$ and decreases with redshift.
A similar trend in $v_{\rm rot}/\sigma_*$ is required to reproduce the
observed evolution of the fraction of galactic bars and of the sSFR discussed
above (as already noted by e.g., Sheth et al. 2008).  As a note of caution
we stress that the above cited studies focus on the gas dynamics instead of
the stellar one, and that the gas component, being subject to additional
forces of radiative and hydrodynamical nature, could have a different dynamics
with respect to the stars. This, together with the very limited number of
galaxies in the samples listed above and the large observational uncertainties
in $v_{\rm rot}/\sigma_{\rm gas}$ prevents us to perform a more quantitative
analysis.

 Recently Kraljic et al. (2012) have studied the occurrence of bars in
  disk galaxies of $10^{10} \lsim M_*/\rm M_{\odot} \lsim 10^{11}$ in 
  high-resolution cosmological simulations. They presented a physically motivated
  scenario in which long-lived bars form when galaxies stop being battered by
  frequent minor mergers, which tend to keep the host galaxies dynamically
  hot (see also Romano-Diaz et al. 2008). In the mass range they studied, the
  bar fraction is $\approx 0$ at $z\gsim 1.5$, $\approx 10 \%$ at $z \approx
  1$, and $\approx 80 \%$ at $z \approx 0.5$, in reasonable agreement with our
  model and with the value of $M_{\rm knee}$ we find for those masses. An
  observational confirmation of the model should pass through an estimate of
  the fraction of galaxies of a given mass undergoing a minor merger within a
  given redshift. This exercise has been already performed in the literature (e.g., Jogee et al. 2009, Lotz
  et al. 2011). The results  depend on the different tracers 
  used to identify galaxy  mergers  and the different assumptions 
  underlying the estimates of the merging frequency, and do not   always agree with the theoretical predictions, 
  calling for a critical   revision of both the observational and theoretical 
  approaches (as discussed   in Lotz et al. 2011). Even so, assuming the maximum 
  number of mergers within   the last 7 Gyr ($z\approx 0.8$) reported in Jogee et 
  al. (2009) and Lotz et al. (2011), about half of the galaxies with $M_* > 10^9 M_{\odot}$ did not
  undergo any minor merger, leaving sufficient time to develop a
  bar in their central regions.

We conclude by commenting that the $M_{\rm knee}\propto (1+z)^2$ fit to the
observational data (Figure~\ref{ssfr}, panel b) implies that 
$v_{\rm rot}/\sigma_*$ does not depend on two uncorrelated variables ($M_*$ and
$z$), but only on $M_*/(1+z)^{2}$, decreasing the dimensionality of the
problem. This prediction can be tested by future accurate measurements of
$v_{\rm rot}/\sigma_*$ in larger samples of galaxies of different masses and
redshifts.

% {\mf In una earlier version c'era un conto che prevedeva il (z+1)$^2$ trend.
% Vista la nota del caveat, si puo' lasciare il conto. Se si decide di toglierlo, 
% pero' dobbiamo aggiornare asbtract di conseguenza, in quanto non facciamo piu' 
% previsione esplicita di redshift dependence... O mi sono perso qualcosa?
% MASSI puoi intervenire tu?}
%
% We check anyway the broad
%consistency of the observed $v_{\rm rot}/\sigma_{\rm gas}$ and $M_{\rm knee}$
%evolutions as a function of $z$, under the simplifying assumption that
%$\sigma_*\approx \sigma_{\rm gas}$. More detailed investigations are postponed
%to future investigations.
%
%The $M_{\rm knee}\propto (1+z)^2$ fit to the observational data
%(figure~\ref{knee}) implies that $v_{\rm rot}/\sigma_*$ can be written as a
%function $f(M_*/(1+z)^{2})$......
%
%
%We make use
%of the collection of observational values of $v_{\rm rot}/\sigma_{\rm gas}$
%presented in Sesana et al. (2014) to estimate of $\bar{M}$ and $c$.  We use
%data only for galaxies with $M_*>10^{9.5}$ M$_{\odot}$, since are the only
%ones showing bars at any observed redshift. The best fit, compared to the
%observational data in figure~\ref{fit}, is obtained for $\bar{M}=xxx$ and
%$c=yyy$.
%
%  
%
%\begin{figure}[!t]
%\centering
%\includegraphics[width=5cm, trim = 0cm 0cm 0cm 0cm]{intercz.eps}
%\caption{
%
% }
%\label{fit}  
%\end{figure}

\section{Discussion}
\label{speculations}

The results of our observational study together with simple numerical and
analytical arguments, demonstrate the relevance of bars in quenching the
central regions of about $25\%$ of the field main-sequence galaxies with
$M_*>M_{\rm knee}$ in our sample. In this section we speculate further,
depicting a physical scenario in which the mass quenching of the vast majority of
the massive field galaxies is caused by the occurrence of a bar.

\begin{figure}[!t]
\centering \includegraphics[width=8cm, trim = 0cm 0cm 0cm 0cm]{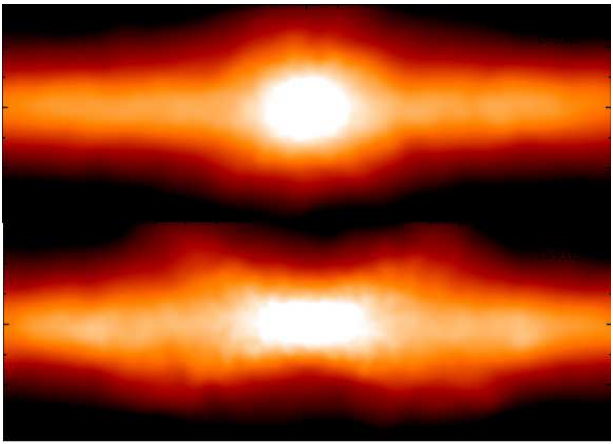}
 \caption{Edge-on view of the same object at the $t$=9 Gyr step of the simulation 
in Figure \ref{simul}. The azimuthal angle of the bar is along (perpendicular to) the line 
of sight in the top (bottom) panel. In both cases we would classify it as pseudobulge.}
 \label{edgeon}  
\end{figure}

In section~\ref{bars} we visually classified the galaxies in our sample 
either as pure disks or as hosts of  strong bars or bulges. A significant fraction
($\sim 40\%$) of the galaxies above $M_{\rm knee}$ do not show a prominent
bar, but rather host a central bulge.  We start by assuming that the most
 of the observed bulges are not classical, but rather boxy/peanut
bulges and/or pseudobulges.
Such an assumption is not unrealistic for isolated disk galaxies (Weinzirl et al. 2009), but
we stress again that it is highly challenging to
classify reliably  the different bulge morphologies (e.g., Graham et al 2008; Wilman et al. 2013). 
The task is even harder as recent evidence indicates that the two bulge categories 
can even occur simultaneously (Erwin et al.  2015).  Given the 
speculative nature of this section, we will work under this assumption anyway,
for which pseudobulges originate from bars. 
In fact, there is a growing evidence (e.g., Combes et al. 1990; Kormendy
\& Cornell 2004; Athanassoula et al. 2005, 2008)
that non-classical bulges represent the late evolutionary stage of stellar
bars, due to the buckling of the central part of the bar itself (boxy/peanut
bulges) and to nuclear star formation fueled by bar-driven gas inflows
(pseudobulges). The formation of a central gas concentration that could result
in the formation of a pseudobulge is present in the simulation previously
discussed (section~\ref{simulations}). We further note that at the very end
of the simulation the bar develops  a thicker rotating stellar
structure in its center, consistent with a boxy/peanut shape  bulge, depending on the assumed
line of sight (see Figure~\ref{edgeon}). Meanwhile, although still present, 
the bar becomes harder to identify in the stellar surface density distribution.
For a detailed theoretical and observational description of the
bar/pseudobulge interplay we refer the reader to the work of 
Raha et al. (1991), Kormendy \& Kennicutt (2004), Kormendy (2013), and Sellwood (2014).

While it is well established that pseudobulges and boxy/peanut bulges can
indeed form from the evolution of a bar, we cannot prove that our bulge
category does not include a significant fraction of classical bulges. A
detailed study of the nature of the bulges would require a wealth of
additional information, including observational constraints on the dynamical
state of the bulges (e.g., through long slit or integral field spectroscopy),
and  is beyond the scope of this paper. However, we can discuss some
additional properties of the bulge population that hint to a physical link
with the population of bar-hosting galaxies. 

Figure~\ref{barre} shows a sharp separation between pure disks and bars or
bulges across $\rm 10^{9.5}$ M$_\odot$. A Kolmogorov-Smirnov test gives a null
probability that the distributions of pure disks and galaxies hosting bars or
bulges are derived from the same parent population.  Bars and bulges, instead,
have almost identical SFR distributions ($>$99\% K-S probability, see the
right panel of Figure~\ref{barre}), supporting a scenario in which bars and
bulges are physically associated. Bars and bulges also show  similar mass
distributions (upper panel in Figure~\ref{barre}). In this last case, however,
they do not perfectly match, probably because of  the ambiguity in the
classification of objects in the transition regime between bulges and disks
near $10^{9.5}$ M$_\odot$.

An additional independent hint comes from the study of the nuclear activity of
galaxies with bars and bulges.  Observations confirm that indeed many barred
galaxies have dense central concentration of gas and enhanced central star
formation (Sakamoto et al. 1999; Laurikainen et al. 2004; Jogee et al. 2005; Sheth et al. 2005;  
Ellison et al. 2011; Kormendy et al. 2013).  We  strengthen
this point by performing an analysis of the nuclear activity of the galaxies in
our sample, making use of the classification given in Gavazzi et
al. (2013c). We find that among massive barred galaxies ($M_* > 10^{9.5}$
M$_\odot$) at $z=0$, 61\% of nuclei show line ratios typical of 
HII regions, 12\% are strong AGNs (mostly
type 2), and 11\% are either passive (2\%) or retired (9\%) \footnote{
``Retired galaxies'' is a denomination proposed by Stasi{\'n}ska et al. (2008) 
to describe nuclei that have stopped forming stars and are ionized by ``hot post-AGB stars''.}.  
The remaining galaxies (15\%) are classified as LINERs.  Very similarly, 
among massive spirals showing bulges, 53\% have HII-like nuclei, 14\% are strong AGNs, 23\%
are LINERs, and 9\% are passive or retired. The large fraction of star-forming
nuclei and strong AGNs in the two samples hints at large gas concentrations,
and the similar fractions further hint to a common physical origin of bulges
and bars. As a check we performed the same exercise among 954 E+S0, selected
in the Local and Coma superclusters with stellar masses greater than
$10^{9.5}$ M$_\odot$. Of these, only 5\%  show line ratios common to HII regions, 
2\% are strong AGNs, while 13\% are LINERs and the remaining 80\% are either passive or
retired. In summary, the population of galaxies with strong bars is indistinguishable 
from that hosting bulges as far as their nuclear properties are concerned,
while the E+S0 class (supposedly dominated by genuine classical bulges) does
not show any significant central activity whatsoever.

In conclusion, the arguments discussed so far support a 
possible evolutionary scenario in  which, at a given redshift,
galaxies above $M_{\rm knee}$ undergo a bar
instability (section~\ref{dynamics}). The bar forces the gas within the
corotational radius to fall toward the center in few dynamical times. The
forming central gas condensation is immediately consumed by a vigorous burst
of star formation (and/or AGN activity), resulting in the formation of a
pseudobulge. After a few rotations, the bar sweeps all the gas within its
corotational radius, quenching the SF in the central region of the
galaxy. Consequently, this region grows redder and redder with time,
decreasing the global sSFR of the galaxy (see also Cheung et al. 2013).  With time the central region of the
bar undergoes a buckling instability (e.g., Sellwood 2014 and references
therein): the bar becomes less and less visible, while a thicker but still
rotationally supported stellar condensation (i.e., a boxy/peanut bulge) becomes
clearly observable, often with a pseudobulge hosted in its very center. The
common origin of pseudobulges and boxy/peanut bulges from bars 
justifies $i)$ the significant fraction of galaxies hosting bulges 
observed in our sample above $M_{\rm knee}$ and $ii)$ the similarities 
between their masses, SFRs, and nuclear activity distributions and those 
describing their barred counterparts.

\section{Summary and conclusion}
\label{discussion}

 In the present paper we tried to reconstruct the star formation history of main-sequence galaxies 
  as a function of stellar mass from the present epoch (Section \ref{z0}) up to $z=3$ (Section \ref{highz}). 
  The local determination was based on the H$\alpha$ narrow-band imaging follow-up
  survey (H$\alpha$3) of field galaxies selected from the HI Arecibo Legacy
  Fast ALFA Survey (ALFALFA) in the Coma and Local superclusters. 
  The higher redshift measurements were taken from the recent literature.
  
  A clear evolutionary trend was found indicating that star-forming galaxies had their star formation rate quenched
  above a certain threshold mass, which is a strong increasing function of redshift (Section \ref{highz}).  
    
  To help identify what physical mechanism is responsible for this mass quenching,
  a set of hydrodynamical simulations of isolated disk galaxies was run to reproduce the formation of a bar (Section \ref{simulations}) and 
  some dynamical considerations allowed us to highlight the joint dependence on mass and redshift of the Toomre conditions for bar instability (Section \ref{dynamics}).
      
The present investigation has focused on five fundamental aspects underlying the
global history of star-forming galaxies:\\ $(i)$ there is a clear increase in
the fraction of visually classified strong bars above some critical stellar mass $M_{\rm knee}$ that in the
local Universe corresponds to $\sim 10^{9.5}$ M$_\odot$; \\ $(ii)$ above
$M_{\rm knee}$ the bars are responsible for intense gas inflows that
effectively trigger bursts of nuclear star formation that accelerate SF
  activity in the circumnuclear region, thus contributing to quenching the
  star formation in the longer run within the bar extent (on kpc
scales)  in agreement with Cheung et al. (2013);\\ $(iii)$ the
critical stellar mass $M_{\rm knee}$ is found to be strongly dependent on
redshift, with only the most massive galaxies harboring bars at high
redshift;\\ $(iv)$ the specific star formation rate below $M_{\rm knee}$
(among normal main-sequence galaxies) strongly increases with redshift at
least up to $z\sim 4$ (Madau et al. 1998);\\ $(v)$ among centrally quenched
galaxies, above $M_{\rm knee}$, the effects of quenching decrease
significantly with increasing redshift.

%  Figure \ref{thumb} provides a vivid illustration of points $(i)$ and $(ii)$:
%  the bar fraction and the consequent regulation of the specific star formation
%  rate (traced by the mean color) depend on stellar mass.   This figure  
%  shows 12 galaxies in our local sample, selected for being nearly face-on to better 
%  discern the presence of bars. The selected objects belong to 4 bins of stellar mass, 
%  two below and two above M$_{knee}$ (top rows).  It is evident that, with increasing mass,
%  the bars  become more conspicuous and contextually the colors become redder,
%  especially within the bar extent.  Vigorous star
%  formation is still taking place in the outer regions of even the most massive
%  spirals.

Points $(iv)$ and $(v)$ may be caused by the cosmic evolution of galaxies,
according to which higher redshift galaxies are progressively more gas-rich
and are more often perturbed. Instead, results $(i)$ to $(iii)$ can be accounted
for within a simple, physically motivated scenario, as detailed in
section~\ref{theory}.  In this picture, galaxies evolve from dynamically
hotter structures to disks clearly dominated by their bulk rotation. More
massive galaxies settle into dynamically cold configurations earlier, as
supported by a growing wealth of observations (see references in
section~\ref{dynamics}), with respect to less massive structures. As soon as a
galaxy relaxes, the central part of the disk can undergo bar instability. The
resulting bar sweeps away the gas within its corotational radius quenching the
SF in the central region of the galaxy. This region, consequently, grows
redder and redder with time, decreasing the global sSFR of the galaxy.

As a note of caution we stress that, although bars play a significant
  role, some additional mass-driven quenching mechanisms are required to explain
  the ``downsizing'' of high-mass spirals. As shown in Figure~\ref{colmag}, even
  the exteriors of massive barred galaxies are redder than lower mass
  counterparts. This could also be related to the evolution of galaxies in a
  cosmological context. Since the additional mechanisms are needed to quench
  the outer regions of field disk galaxies, we consider cosmological
  starvation (Feldmann \& Mayer 2015, Fiacconi et al. 2015, Peng et al. 2015) to be a better candidate
  than    SF/AGN feedback, for example,  or any environmental effect. A complete
  understanding of this second quenching mechanism would require a more comprehensive
  study and  is beyond the scope of this investigation.

The simple model outlined above has a number of testable assumptions and
predictions:  $(i)$ Deep imaging can verify whether the central regions of
quenched galaxies host bars/bulges at higher redshift, and if such structures
are instead absent below $M_{\rm knee}$. This is already hinted at by
observational studies of the cosmic evolution of the bar occupation fraction,
e.g., Sheth et al. (2008). $(ii)$ Our model predicts that the degree of
``relaxation'' of galaxies, as described by the $v_{\rm rot}/\sigma_*$ ratio,
must depend on a specific combination ($M_*/(1+z)^{2}$) of the galaxy masses
and redshift. Increasing the statistics and the accuracy of $v_{\rm
 rot}/\sigma_*$ measurements in mass and redshift bins will test such a
prediction.

We conclude by speculating on the relevance of the bar-induced
  mass quenching for massive field galaxies. We believe that most
  of the massive galaxies that do not show a clear bar while hosting a central
  bulge can be associated with a late evolutionary stage of a previously barred
  galaxy.  In this scenario most of the bulges in our classification would be
  either pseudobulges, formed during the bar-induced nuclear gas inflow, or
  boxy/peanut bulges,  which are the results of the buckling instability
  that naturally develops in the central regions of the bar (e.g., Sellwood 2014 and references
  therein).  As discussed in the literature, the bar buckling and
  formation of dense nuclear concentration of mass (e.g., the pseudobulge)
  modifies the dynamics of the stars in the bar. This can result in what is known as ``bar suicide'': 
  the bar becomes less and less visible (Raha et
  al. 1991; Norman et al. 1996; Martinez-Valpuesta \& Shlosman 2004; Shen \&
  Sellwood 2004; Debattista et al. 2004, 2006; Athanassoula et al.
  2005). A thicker but still rotationally supported stellar condensation
  (i.e., a boxy/peanut bulge) with a pseudobulge hosted in its very center
  would be the remaining traces of the dissolved bar. Such a speculative
  scenario is supported the similarities between the mass, SFR, and nuclear
  activity distributions of massive galaxies hosting bars and bulges.

 %\newpage
\begin{acknowledgements}
We thank Katherine Whitaker for exchanging information with us, Davide
Fiacconi for his help in initializing the numerical simulation presented in
this work, and Alister Graham for comments.  We wish to thank the anonymous
referee for his constructive criticism.
The authors would like to acknowledge the work of the entire
ALFALFA collaboration team in observing, flagging, and extracting the catalog
of galaxies used in this work and thank Shan Huang for providing original
data.  This research has made use of the GOLDmine database (Gavazzi et
al. 2003, 2014b) and of the NASA/IPAC Extragalactic Database (NED) which is
operated by the Jet Propulsion Laboratory, California Institute of Technology,
under contract with the National Aeronautics and Space Administration.  We
wish to thank an unknown referee whose criticism helped improving the
manuscript.  Funding for the Sloan Digital Sky Survey (SDSS) and SDSS-II has
been provided by the Alfred P. Sloan Foundation, the Participating
Institutions, the National Science Foundation, the U.S. Department of Energy,
the National Aeronautics and Space Administration, the Japanese
Monbukagakusho, and the Max Planck Society, and the Higher Education Funding
Council for England.  The SDSS Web site is \emph{http://www.sdss.org/}.  The
SDSS is managed by the Astrophysical Research Consortium (ARC) for the
Participating Institutions.  The Participating Institutions are the American
Museum of Natural History, Astrophysical Institute Potsdam, University of
Basel, University of Cambridge, Case Western Reserve University, The
University of Chicago, Drexel University, Fermilab, the Institute for Advanced
Study, the Japan Participation Group, The Johns Hopkins University, the Joint
Institute for Nuclear Astrophysics, the Kavli Institute for Particle
Astrophysics and Cosmology, the Korean Scientist Group, the Chinese Academy of
Sciences (LAMOST), Los Alamos National Laboratory, the Max-Planck-Institute
for Astronomy (MPIA), the Max-Planck-Institute for Astrophysics (MPA), New
Mexico State University, Ohio State University, University of Pittsburgh,
University of Portsmouth, Princeton University, the United States Naval
Observatory, and the University of Washington.\\ 
M. Fossati acknowledges the support of the Deutsche Forschungsgemeinschaft via Project ID 387/1-1.
R.G. and M.P.H. are supported by US NSF grants  AST-1107390 and the Brinson
Foundation grant.
M. Fumagalli acknowledges support by the Science and Technology Facilities Council 
[grant number ST/L00075X/1]. H. Hernandez Toledo acknowledges support from DGAPA PAPIIT grant IN112912.
L. Gutierrez acknowledges support from Consejo Nacional de Ciencia y Tecnología de México (CONACYT) under project 167236.

\end{acknowledgements}

%\newpage
\end{document}